\documentclass[traditabstract]{aa} 
\usepackage{graphicx,longtable,cancel}
\usepackage{txfonts}
\begin{document}
\bibliographystyle{aa}
% Units                                                                                                                                  
\newcommand{\um}{$\mu$m}
\newcommand{\uJy}{$\mu$Jy}
\newcommand{\ergs}{{erg\,s$^{-1}$}}
\newcommand{\ergscm}{{erg\,s$^{-1}$\,cm$^{-2}$}}
\newcommand{\kms}{{km\,s$^{-1}$}}
\newcommand{\msun}{${\rm M}_{\odot}$}
\newcommand{\lsun}{${\rm L}_{\odot}$}
\newcommand{\msunyr}{${\rm M}_{\odot}~{\rm yr}^{-1}$}
\newcommand{\lir}{$L_{\rm IR}$}
\newcommand{\Lbol}{$L_{\rm bol}$}
\newcommand{\ms}{$M_{\rm star}$}
\newcommand{\mg}{$M_{\rm gas}$}
\newcommand{\mh}{$M_{\rm halo}$}
\newcommand{\mbh}{$M_{\rm BH}$}
\newcommand{\mui}{$\mu^{-1}$}
\newcommand{\cohh}{CO$\to$H$_2$}
\def\Zsun{Z_\odot}
\def\gtsima{$\; \buildrel > \over \sim \;$} 
\def\ltsima{$\; \buildrel < \over \sim \;$} \def\prosima{$\; \buildrel \propto \over \sim \;$} 
\def\gsim{\lower.5ex\hbox{\gtsima}}
\def\simgt{\lower.5ex\hbox{\gtsima}} 
\def\simlt{\lower.5ex\hbox{\ltsima}} 
\def\lsim{\lower.5ex\hbox{\ltsima}}
\def\msole{M_\odot}
\def\rsole{R_\odot}
\def\msun{M_\odot}
\def\rsun{R_\odot}
\def\mdot{\dot{M}}
\def\OVI{\hbox{O~$\scriptstyle\rm VI$}}
%units,expressions                                                                                                                       
\def\invMpch {~h~{\rm Mpc}^{-1}}
\def\Mpch {~h^{-1}~{\rm Mpc}}
\def\kpch {~h^{-1}~{\rm kpc}}
\def\swift {{\it Swift}}
\def\xmm {{\it XMM-Newton}}
\def\chandra {{\it Chandra}}
\makeatletter
\def\@biblabel#1{\hspace*{-\labelsep}}
\makeatother

\title{Multiple tidal disruption flares in the active galaxy IC 3599}

\author{
S. Campana\inst{1}, 
D. Mainetti\inst{1,2}, 
M. Colpi\inst{2}, 
G. Lodato\inst{3}, 
P. D'Avanzo,\inst{1}, 
P. A. Evans\inst{4},  
A. Moretti\inst{1}
}

\institute{INAF, Osservatorio Astronomico di Brera, Via E. Bianchi 46, 23807, Merate (Lc), Italy\\
 \email{sergio.campana@brera.inaf.it}
\and
Dipartimento di Fisica G. Occhialini, Universit\`a degli Studi di Milano Bicocca, Piazza della Scienza 3, I-20126 Milano, Italy
\and
Dipartimento di Fisica, Universit\`a degli Studi di Milano, Via Celoria 16, I-20133 Milano, Italy
\and
Department of Physics and Astronomy, University of Leicester, University Road, Leicester LE1 7RH, UK
              }

\date{}

\abstract{
Tidal disruption events occur when a star passes too close to a massive black hole and it is totally ripped apart by tidal forces. 
It may also happen that the star is not close enough to the black hole to be totally disrupted and a less dramatic event might happen.
If the stellar orbit is bound and highly eccentric, just like some stars in the centre of our own Galaxy, repeated flares should occur.
When the star approaches the black hole tidal radius  at periastron, matter might be stripped 
resulting in lower intensity outbursts recurring once every orbital period.
We report on {\it Swift} observations of a recent bright flare from the galaxy IC 3599 hosting a middle-weight 
black hole, where a possible tidal disruption event was observed in the early 1990s.
By light curve modelling and spectral fitting  we can consistently account for the events
as the non-disruptive tidal stripping of a star into a highly eccentric orbit. The recurrence time is  9.5 yr.
IC 3599 is also known to host a low-luminosity active galactic nucleus. Tidal
stripping from this star over several orbital passages might be able to spoon-feed
also this activity.
}


\authorrunning{Campana et al.}

\maketitle

\section{Introduction}

Supermassive black holes reside at the centre of most galaxies. When accreting matter, these black holes emit a huge amount of 
energy, becoming active galactic nuclei (AGNs). The energy and radiation produced by matter accretion onto these black holes 
play an important role in determining their masses and spins as well as the properties of the host galaxy bulges.
%, through what is known as AGN feedback. 
AGNs usually have a relatively short duty cycle and are in a 
low-luminosity state for most of the time (Ho 2008). AGNs may become bright when major inflows of gas feed the  
compact object (Hopkins et al. 2006). Apart from intense episodes of accretion, the black hole is powered at a floor minimum 
level mostly by a hot gas from colliding stellar winds in the central region. 
A star in the nuclear cluster can be scattered close to the black hole via dynamical encounters and 
put into an orbit too close to the black hole. When the black hole tidal force overcomes the star self gravity (at the so
called tidal radius) the star starts to be ripped apart. About half of the star mass remains bound to the black hole, forming an accretion disc 
and powering a luminous, long lasting (months to years) accretion flare (e.g. Rees 1998; Phinney 1989).
Despite their elusiveness, a number of events have been reported in the literature, 
resulting in the complete disruption of the star (e.g. Renzini et al. 1995; Bloom et al. 2011; Gezari et al. 2012). 

Tidal disruption events (TDEs) provide an unique probe to reveal the presence of otherwise quiescent 
black holes, allowing us to study the accretion process also in a different regime than that of AGNs (Komossa 2012; Gezari  2012). 
TDEs were first predicted and later on observed in the soft X--ray to ultraviolet bands where the 
peak of the accretion disc emission lies (Strubbe \& Quataert 2009). Two relativistic TDEs were discovered based on high energy triggers,
thanks to the launching of relativistic jets along the line of sight (Bloom et al. 2011; Burrows et al. 2011; Cenko et al. 2012).
Only a small number of TDEs  however 
have been observed to date in X--ray, UV and optical surveys, mainly because of their low rate of occurrence and sparse observations 
(Donley et al. 2002).

TDEs are ultimate events. A star entering the tidal radius $r_{\rm t}\sim R_*\,(M_{\rm BH}/M_*)^{1/3}$ (where $M_{\rm BH}$ is the 
black hole mass and $M_*$ and $R_*$ are the mass and radius of the star) is completely disrupted
by tidal forces. The fate of a star orbiting its central black hole is then defined by its pericenter radius, $r_{\rm p}$:
if $\beta=r_{\rm t}/r_{\rm p}\gsim 1$ we have a TDE. What happens instead if $\beta\lsim 1$? If $\beta$ is still close to unity, 
the star should still feel the black hole tidal force but will survive the encounter. 
Hydrodynamical simulations show that if the passage is close ($0.5\lsim \beta \lsim 1$) 
the star may lose matter that can then accrete onto the black hole (Guillochon \& Ramirez Ruiz 2013). 
These events should be more frequent than classical TDEs (roughly by a factor of $\sim 10$) and should repeat on the 
star orbital period (if the star is on a bound orbit), increasing their observability. On the other side, these events are less energetic, 
because the involved mass captured by the black hole is lower.

Here, after a brief introduction on total and partial tidal disruption events (Section 2),  we consider the case of a nearby low-luminosity AGN, IC 3599, 
that showed in the past strong X--ray activity, possibly resulting from a TDE (Brandt, Pounds \& Fink 1995; Grupe et al. 1995, see Section 3). 
In Section 4 we describe new {\it Swift}/XRT data showing a second, large increase in flux, together with the analysis of all X--ray data on the source.  
X--ray spectral analysis is presented in Section 5. Based on spectral results, we converted count rates into a flux and then in luminosity. The fit of the 
overall, 24 yr long, X--ray light curve is described in Section 6. 
Optical data are discussed in Section 7.
In Section 8 we apply partial TDE modelling to the light curve of IC 3599 trying to constrain 
the involved star and its orbit. Our conclusions are in Section 9.

\section{Tidal stripping essentials}

A star orbiting too close to a massive black hole will be torn apart by the compact object tidal force.
This occurs at the tidal radius beyond which stellar self-gravity is not able to counteract the black hole 
tidal field and keep the star together. The tidal radius is defined as
$$
r_{\rm t}\sim R_*\,(M_{\rm BH}/M_*)^{1/3}\sim 23 \,(M_{\bullet}^{\rm IC})^{-2/3} \, r_{\rm g}
$$
where $R_*$ and $M_*$ are the radius and mass of the star,  $M_{\rm BH}$ the mass of the black hole 
and $M_{\bullet}^{\rm IC}\sim 3\times 10^6\msole$ the mass of the central black hole in IC 3599 (Grupe, Komossa \& Saxton 2015, see also below). 
The numerical value refers to a $1\msole$ and $1\rsole$ star and units are gravitational radii ($r_{\rm g}$).

As the star is ripped apart by the tidal forces of the black hole, the debris is thrown into high-eccentricity orbits 
with a large range of periods and with an energy range of (Lacy, Townes \& Hollenbach 1982):
$$
\Delta E\sim {{G\,M_{\rm BH}\,R_*}\over{r_{\rm t}^2}}
$$
The distribution of mass as a function of energy is nearly 
constant (Rees 1988), as also shown by numerical simulations (Evans \& Kochanek 1989; Lodato, King \& Pringle 2009).

For a parabolic orbit, nearly half of the debris is unbound leaving the system at high velocity. The other half will return to the 
black hole at different times depending on the initial eccentricity (i.e. energy). The return time of the first debris at pericentre is
\begin{eqnarray}
t_{\rm min} &=& {{2\,\pi\,G\,M_{\rm BH}}\over{(2\,\Delta E)^{3/2}}} \sim  {{2\,\pi\,r_{\rm t}^3}\over{(G\,M_{\rm BH})^{1/2}\,(2\,R_*)^{3/2}}} = \nonumber \\
                    &=& 71\,R_*^{3/2}\,M_*^{-1}\,(M_{\bullet}^{\rm IC})^{1/2}\ {\rm d} \nonumber  
\end{eqnarray}
The return of material at pericentre continues at a rate driven by Kepler's third law as
$$
\mdot\sim {1 \over 3}\, {M_* \over t_{\rm min}} \, \left( {t \over t_{\rm min}} \right)^{-5/3}
$$
where the maximum mass inflow rate is 
$$
\mdot_{\rm peak}={1 \over 3}\, {M_* \over t_{\rm min}}\sim 1.7\,R_*^{-3/2}\,M_*^2\,(M_{\bullet}^{\rm IC})^{-1/2}\msole {\rm yr}^{-1}
$$
The peak rate is a function of the stellar structure (Lodato et al. 2009).
After disruption, the stellar debris are launched into very eccentric orbits and gradually return to pericentre, where they 
circularise and form an accretion disc at $r_{\rm circ} \sim 2\,r_{\rm p}$. The following fall back of this material onto the black hole 
is governed by viscous times.
If the viscous time is short in comparison with $t_{\rm min}$, then the fall back of matter onto the central object is 
almost instantaneous and $t_{\rm fb}\sim t_{\rm min}$. 
Based on an $\alpha-$ viscosity disc prescription we can evaluate the viscous time-scale $t_{\nu}$ 
$$
t_{\nu}={\frac{t_{\rm Kep}(2\,r_{\rm p})}{\pi\, \alpha\,h^{2}}}
$$
where $t_{\rm Kep}$ is the Keplerian time at a given radius and $h$ is the disc half-height divided by the radius.
Being the initial luminosity close to (or even somewhat larger than) the Eddington limit, the disc is expected to 
be thick (Ulmer 1999), with $h\sim1$.
Thus, we can estimate the ratio of the viscous time to the $t_{\rm min}$ time as
$$
{{t_{\nu}} \over {t_{\rm min}} } \sim 1.8\times 10^{-3}\,\beta^{-3/2}\,\left(\frac{\alpha}{10^{-1}}\right)^{-1}\,(M_{\bullet}^{\rm IC})^{-1/2}\,M_{*}^{1/2}\,h^{-2}
$$
and a thick disc will form and drain as the material circularises down to the last stable orbit.

The infalling star is usually assumed to be in a parabolic orbit and to undergo complete tidal disruption.
Encounters may still happen with the pericentre slightly outside the tidal radius ($\beta\lsim1$),  resulting in some spill over of stellar 
matter, and orbits can be bound and highly eccentric, resulting in periodic outbursts.
This problem received less attention than the classical TDEs but still a number of works 
exist (Guillochon \& Ramirez Ruiz  2013;  Macleod, Guillochon \& Ramirez-Ruiz 2012;  
Macleod et al. 2013; Hayasaki, Stone \& Loeb 2013; Ivanov \& Novikov 2001).
If the orbit is bounded the physics is similar to the parabolic case in terms of the effects of black hole tides on the 
approaching star, especially if the eccentricity is high.
On the long term the star orbit may change but the typical orbital binding energy is larger than the stellar binding energy 
and the orbital angular momentum is larger than the rotational angular momentum of the star, indicating that the transfer of mass
will not substantially alter the orbit (Macleod et al. 2013).

In particular, Guillochon \& Ramirez Ruiz (2013) investigated the fate of a star undergoing partial stellar stripping.
They considered the case of stellar polytropes of $\gamma=4/3$ and $\gamma=5/3$ and investigated the case for $\beta$ 
within the interval 0.6--4.0 and 0.5--2.5, respectively.
They found that total disruption occurs for a critical impact parameter different from unity, being $\beta_c=1.85$ for $\gamma=4/3$ and  
$\beta_c=0.90$ for $\gamma=5/3$, respectively. They also provided fitting formulae for estimating the mass accretion peak rate,
the peak time, and the total mass transferred for both cases, which are strong functions of $\beta$, especially for $\beta \lsim \beta_c$ 
(see also the related errata corrige at http://astrocrash.net/2013/09/16/errata-of-guillochon-ramirez-ruiz-2013/). 
We used these formulae in deriving stellar parameters.

\section{IC 3599}

One of the first putative TDEs occurred in the close active galaxy IC 3599 (92 Mpc, at a redshift of $z=0.021$; Brandt et al. 1995; Grupe et al. 1995). 
It was discovered in the X--ray band as a bright, soft source during the {\it ROSAT} all sky survey in Dec. 1990. 
Further {\it ROSAT} observations (June 1992 to June 1993) found IC 3599 in a dimmer ($\sim 100$) and somewhat spectrally 
harder state (see Fig. 1). A {\it Chandra} observation in March 2002 found the source at a similar level (Vaughan et al. 2004).
The very large flux decrease and the softness of the X--ray spectrum led several authors to suggest that  this outburst was a 
TDE (Brandt et al. 1995; Grupe et al. 1995; Komossa \& Bade 1999; Vaughan, Edelson \& Warwick 2004), even if IC 3599 is an active galaxy. 
Indeed active galaxies are predicted to host more TDEs due to the perturbing
presence of the disc, even if it would be more difficult to reveal
them because of the higher overall emission (Komossa 2012; Karas \& Subr 2007). 
The optical spectrum is characterised by strong H$\alpha$, H$\beta$ and [OIII] lines, 
showing variations in response to the X--ray outburst and classifying IC 3599 as a type 1.5-1.9 Seyfert 
galaxy (Brandt et al. 1995; Grupe et al. 1995; Komossa \& Bade 1999). 
%Super-Eddington tidal flares can produce photoionised emission lines in the outflow as well as 
%low velocity absorption lines (Strubbe \& Quataert 2011), possibly explaining the differences in the classification. 
The central black hole mass has been derived from a relation between the flux at 5100\,\AA\ and the width of the quiescent H$\beta$ 
emission line to be $M_{\bullet}^{\rm IC}\sim 3\times 10^5\msole$ (Sani et al. 2010).
Grupe et al. (2015) argued that this estimate has been obtained using a Broad Line Region scaling relation, which is appropriate 
only in case of an unabsorbed line of sight. Using instead a black hole mass to bulge $K$-band luminosity (Marconi \& Hunt 2003) or the relation
between the [OIII] velocity dispersion and the black hole mass (Nelson 2000), they found
$M_{\bullet}^{\rm IC}$ to be in the $(2-12)\times 10^6\msole$ range.
Here we adopt a black hole mass of $3\times10^6\msole$. 
For this mass the source was well below the Eddington limit during the {\it ROSAT} all sky survey observation.

The (low-luminosity) AGN nature of IC3599 may cast doubts on the tidal event interpretation, because AGNs show 
flares, which are related to disc activity or to the uncovering of a heavily absorbed X--ray source.
We note here that the quiescent {\it Chandra} spectrum is well described by a soft power law with minimal intrinsic 
absorption, weakening considerably the absorbed source case. 
At the same time variations by a factor of $\gsim 20-30$ in AGNs are very rarely observed. 
We study in details AGN variability and, based on a sample of highly variable AGNs intensively monitored by {\it Swift}, we can assess that 
a variability similar to the one observed in IC 3599 occurs by chance at $\sim 4.5\,\sigma$ level (see Appendix A).  In particular,
the second flare should result by chance from known AGN variability at $\gsim 4.0\,\sigma$. 
Other models apart from AGN variability, have been proposed but should not work in the case of IC 3599.
A binary system made by the central black hole and an orbiting star filling its Roche lobe at periastron can be ruled out as the stellar 
density required would be too low (Lasota et al. 2011). In addition, we confirm that the X--ray and optical transient position is consistent with the 
centre of IC 3599 (see Appendix B).
Periodic optical outbursts were observed in the BL Lac object OJ 287 and explained by 
accretion instabilities onto a binary black hole (Tanaka 2013). However, this mechanism does predict no X--rays in quiescence because the 
innermost part of the disc is depleted.
AGN instabilities in slim disk has also been put forward to explain this variability (Honma, Matsumoto \& Kato 1991).
However, the estimated mass accretion rate for IC 3599 falls well below the instability region and the duration of the flares is much 
longer than theory predictions (Xue et al. 2011, see Appendix).  
All these findings lend support to the idea that IC 3599 underwent TDEs.
Two different TDEs appear however unlikely. The occurrence probability of a TDE in a galaxy is of the order of
$10^{-5}$ galaxy$^{-1}$ yr$^{-1}$ (e.g. Donley et al. 2002), so having two different events in 25 yr results in a probability of $\sim 6\times 10^{-8}$.
Even increasing the rate by a factor of 100 as a result of central binary black hole merger (e.g. Perets, Hopman \& Alexander 2007),
would result in a still low probability $\sim 6\times 10^{-4}$.
A binary disrupted by the central black hole would result into two different events but the time delay between them is way too short (several days) 
to account for what we observe (Mendel \& Levin 2015).

%Apart from the strong outburst, the `quiescent'  X--ray flux of IC 3599 shows (if any) only small ($\sim 2$) variability  (see Fig. 2). 
%All these findings lend support to the idea that  IC 3599 underwent a TDE.
IC 3599 was also detected in the radio band by the Very Large Array on June 2012. 
This radio observation is difficult to be accounted for challenging the prediction of the standard blast wave model (Bower et al. 2013).

\section{X-ray data preparation}

X--ray data were collected over a time basis of 24 years with different satellites. A log of all the observations is shown in Table 1.

\subsection{{\it ROSAT} All Sky Survey data}
RASS data are not straightforward to analyse. They were collected using the PSPC detector in scan mode.
For this reason we decided to stick to the original analysis by Grupe et al. (1995).

\subsection{{\it ROSAT} PSPC}
Data were extracted using the task {\tt XSELECT} (v. 2.4c) from a circular region centred on source with a 60 arcsec radius.
The background spectra were extracted from an annular region centred on IC 3599, free of contaminating sources and support shadows, 
with inner and outer radii 125 and 250 arcsec, respectively. Data were retained in the 12--211 channel range, corresponding to 
an energy range of 0.1--2.4 keV. 
Being IC 3599 always on-axis we used the pre-canned response matrix relative to PSPC-B, gain-2 period (pspcb\_gain2\_256.rsp).
Given the relatively low number of photons we binned the data to a minimum of one photon per energy channel
using the {\tt grppha} tool and adopted Cash-statistics to fit the data. 
Different observations were grouped together in order to increase the number of photons as indicated in Table 1.

Cash-statistics does not allow a reliable estimate of the goodness of fit. We estimate the goodness of fit by finding the best fit with 
Cash-statistics and then test the data with $\chi^2$ statistics, corrected for the low number of photons per bin through the 
Churazov weighting scheme (Churazov et al. 1996). This allows us to determine the goodness of fit using $\chi^2$ statistics.

\subsection{{\it ROSAT} HRI}
Data were first combined into a single event file using {\tt XSELECT}. Source photons were extracted from a circular region of 
15 arsec radius. Background photons were extracted from a nearby circular region of 45 arcsec radius.
No spectral data can be obtained from {\it ROSAT} HRI data and the count rate was converted into a {\it ROSAT} PSPC count rate
using PIMMS (see below).

\subsection{{\it Chandra} ACIS-S}
Data were reprocessed using the CIAO 4.6 (and CALDB 4.5.9) {\tt repro} task. 
Spectral data were extracted using the CIAO task {\tt specextract} using a circular 3 arcsec region for the source and an annular region
for the background with 12 and 17 inner and outer radii, respectively. The task provides the user with the corresponding response matrix (rmf) and 
ancillary response file (arf).
Data were retained in the 0.3--10 keV energy range and were binned to a minimum of one photon per energy channel 
using the {\tt grppha} tool.

\subsection{{\it XMM-Newton} Slew survey}
The field of IC 3599 was observed during one XMM-Newton slew on 2007 June 19 (MJD 54270). 
The source was not detected with a ($2\,\sigma$) upper limit of 0.8 counts s$^{-1}$ in the EPIC-pn instrument (Grupe et al. 2015). 
This implies an upper limits on the 0.3--10 keV unabsorbed flux of $1.4\times10^{-12}$ erg cm$^{-2}$ s$^{-1}$ assuming the quiescent 
spectral model (see below). We do not consider this observation any further.

\subsection{{\it Swift} XRT}
Data were reprocessed using {\tt xrtpipeline} (v. 0.12.9) (CALDB 2014-07-30). 
%In the first observation part of the IC 3599 spot falls below bad columns limiting the 
%observed count rate to $\sim 0.2$ c s$^{-1}$ and posing no pile-up problems. In all the other observations the rate is even lower.
Data were extracted from a circular radius of 71 arcsec and the background from a nearby circular region of 141 arcsec radius.
Data were retained in the 0.3--10 keV energy range for spectroscopy and were binned to a minimum of one photon per energy channel
using the {\tt grppha} tool.

\begin{table*}[!tb]
\caption{IC 3599 X--ray observation log.}
\begin{center}
\begin{tabular}{cccccc}
\hline
\hline
Instrument        & Obs. ID.         & Obs. start             & Duration (ks) & Counts$^*$ & Spectrum\\
\hline
{\it ROSAT} RASS & rs931231n00 & 1990-12-10 & 0.8  & --                    & -- \\ 
{\it ROSAT} PSPC& rp700552a00& 1991-12-15 & 3.3  & $101\pm11$ & 1 \\
{\it ROSAT} PSPC& rp700552a01& 1992-05-31 & 1.4  & $33\pm6$      & 2 \\
{\it ROSAT} PSPC& rp701098n00 & 1992-06-16 & 1.8 & $36\pm6$      & 2 \\ 
{\it ROSAT} PSPC& rp701099n00 & 1992-06-17 & 1.9 & $33\pm6$      & 2 \\
{\it ROSAT} PSPC& rp701100n00 & 1992-06-18 & 2.0 & $36\pm6$      & 2 \\
{\it ROSAT} PSPC& rp701097n00 & 1992-06-30 & 5.4 & $116\pm12$ & 3 \\
{\it ROSAT} PSPC& rp701528n00 & 1993-06-17 & 3.4 & $26\pm5$      & {\it 3}\\
{\it ROSAT} HRI    & rh702704n00 & 1996-06-30 & 14.7 & $77\pm11$ & {\it 3}\\
{\it ROSAT} HRI    & rh702706n00 & 1996-06-30 & 16.9 & $87\pm12$  & {\it 3}\\
{\it Chandra} ACIS-S & acisf029999N003 & 2002-03-07 & 10.6 & $427\pm21$ & 4 \\
{\it Swift} XRT         & 00037507001 & 2010-02-25 & 2.1 & $432\pm23$ & 5 \\
{\it Swift} XRT         & 00037507003 & 2010-05-17 & 1.2 & $100\pm11$  & 6\\
{\it Swift} XRT         & 00037507004 & 2013-10-30 & 4.8 & $21\pm6$      & 7 \\
{\it Swift} XRT         & 00037507005 & 2013-11-06 & 4.9 & $12\pm4$      & 7 \\
{\it Swift} XRT         & 00037507006 & 2014-03-26 & 4.7 & $14\pm4$      & 7 \\
{\it Swift} XRT         & 00037569001 & 2014-08-08 & 1.6 & $<16$             & -- \\
{\it Swift} XRT         & 00037569001 & 2014-11-15 & 4.6 & $18\pm5$      & 7 \\
{\it Swift} XRT         & 00037569001 & 2014-11-23 & 0.4 & $<38$             & -- \\
\hline
\hline
\end{tabular}
\end{center}
\medskip
\noindent $^*$ The total number of counts was determined using the XIMAGE (v. 4.5.1) task {\tt detect} (correcting for vignetting 
and point spread function losses).\\
\noindent Spectra were grouped according to the scheme here reported: spectra with the same number were grouped together. 
Numbers in italic means that we used that given conversion factor to pass from count rates to fluxes even if no spectra analysis 
were carried out on these data due to the too low number of counts. \\
\noindent Two {\it ROSAT} PSPC observations on 1992-06-30 and 1992-12-08 were not considered being IC 3599 too 
far off-axis to be detected.\\
\label{obslog}
\end{table*}

\begin{figure}[!ht]
\centerline{
\includegraphics[width=0.45\textwidth,angle=-90]{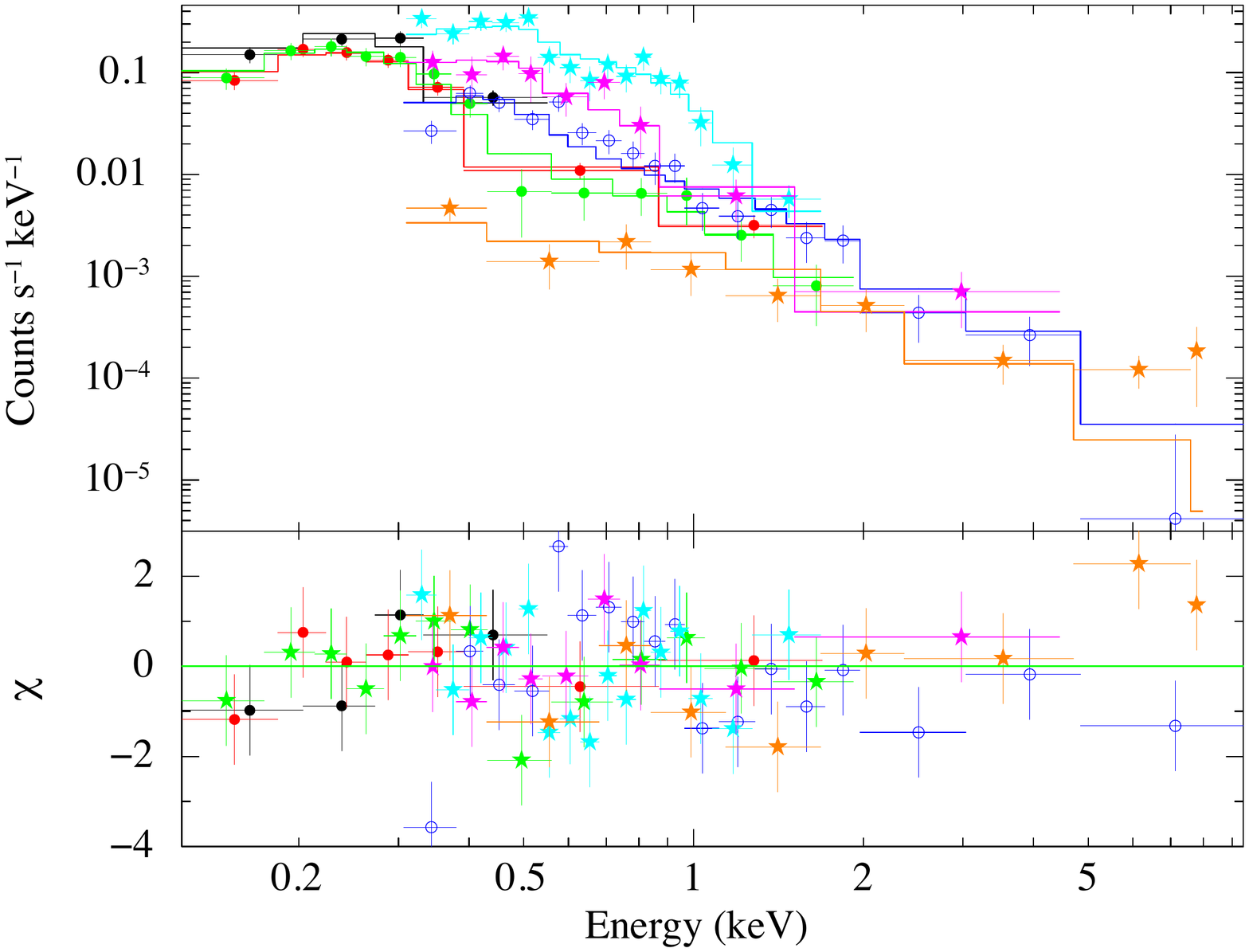}}
\caption{{X--ray spectra fitted with the absorbed disc black body plus power law model described in the text. Data were rebinned to reach 
a $5\,\sigma$ significance or by a factor of 5 for plotting purposes only. In the upper panel black, red and green (filled 
circles) data refer to {\it ROSAT} spectra (first, second and third observations, respectively). Blue (open circles) data refer to the 
{\it Chandra} spectrum. Light blue, magenta and orange (filled stars) to {\it Swift} spectra (first, second and third observations, respectively).
In the lower panel residuals in terms of $\chi$ are plotted with the same colour codings and symbols.
}}
\label{fig1}
\end{figure}

\section{X--ray spectral analysis}

We fitted together three {\it ROSAT} PSPC, one {\it Chandra} ACIS-S and three {\it Swift} XRT spectra (see Table 2) with the X--ray spectral
fitting package XSPEC (v. 12.8.1g).
All spectral fits were minimised using C-statistics and the goodness of the fit was assessed using the Churazov-weighted $\chi^2$ statistics. 
It is readily apparent that the overall X--ray spectra are soft, as testified by previous X--ray data analysis (Brandt et al. 1995; Grupe et al. 1995;
Vaughan et al. 2004).
An absorbed (using {\tt TBABS}) power law model  with all the column densities tied together and the same photon index for all the observations
returns a $\chi^2-$statistic value of 730.9 with 410 degrees of freedom. This power law fit provides a null hypothesis probability 
of $10^{-20}$. The power law photon index is very soft $\Gamma=3.9\pm0.2$ (all errors were determined for $\Delta C=1.0$,
i.e. $1\,\sigma$ errors for one parameter of interest).
Leaving free the power law photon index in each spectrum improves the fit. We obtained $\chi^2=493.3$ with 404 degrees of freedom. 
The corresponding null hypothesis probability is still $0.2\%$. In addition, the photon index of the first {\it ROSAT} spectrum is extremely high 
with $\Gamma=5.0^{+0.4}_{-0.4}$ and the column density $N_H=(4.4\pm0.1)\times 10^{20}$ cm$^{-2}$ is much larger than 
the Galactic column density of $N_H=1.2\times 10^{20}$ cm$^{-2}$ (Kalberla et al. 2005). 
A free column density model with a power law photon index fixed instead provides $\chi^2=619.4$ with 404 degrees of freedom, corresponding to 
a null hypothesis probability is $10^{-11}$. 
A soft model such as a single black body provides similar results: $\chi^2=551.7$ with 404 degrees of freedom, with free temperature and radius.
The null hypothesis probability is $10^{-6}$. It is apparent from the fit that the black body model fails to account for a high energy tail ($>2 $ keV) 
at low fluxes.

We also considered a two component model. Given the large number of possibilities we explored two different models, based on the 
physics of possible emission mechanisms.
The first model comes from the proposal that this bursting activity comes from the partial disclosing of a heavily absorbed AGN. We thus model the 
X--ray spectra with a Galactic absorption plus a partial covering of a power law component. 
The second model is physically motivated by a tidal disruption event. It consists of an accretion disc spectrum (accounting for the disc emission caused by 
the tidal spoon-feeding) and a power law accounting for the the AGN activity (MacLeod et al. 2013).

Spectral fitting results are shown in Table 2.
We explored a partial covering factor model ({\tt pcfabs} within XSPEC) trying to see if the outburst observed in IC 3599 can be 
explained as the unveiling of a
heavily absorbed source. This model has been envisaged to explain the wild erratic variations observed in WPVS 007 (Grupe et al. 2013). 
The fit is carried out with a model made by a fixed Galactic column density plus a partial covering model with the equal intrinsic 
column densities tied to all the spectra but variable fractions. This composite absorption component
screens a power law component with the same photon index but variable normalisation. 
We obtained a $\chi^2=627.4$ for 304 degrees of freedom, resulting in a null hypothesis probability of $10^{-24}$ (see Table 2). 

In the second model parameters are tied among different spectra. The absorbing column density is considered equal for all the spectra.
The {\tt diskbb} temperature is expected to vary, whereas its normalisation was tied between the spectra (being related to the disc inner 
radius  and, likely, corresponding to the innermost stable orbit).
The power law photon index was tied between spectra but its normalisation was free to vary.
The overall fit provides a $\chi^2=315.9$ for 303 degrees of freedom. The null hypothesis probability is $29\%$. 
Residuals are well distributed all over the entire 
energy band (see Fig. 1 and Table 2). The derived column density is somewhat in excess of the Galactic value.
From the normalisation of the disc black body model one can estimate the inner disc radius 
(modulo the disc inclination, $i$, and a colour factor uncertain by a factor of $f_c\sim 2$). 
Considering a maximally rotating black hole, we can estimate it mass as  $\sim 
8\times 10^5\times (\cos{i}/2)^{1/2} \times (f_c/2)\msole$.
This estimate, even if approximate, is barely consistent with the value derived from optical studies,  indicating that the 
central black hole in IC 3599 is not particularly massive.

\begin{table}[!htb]
\caption{IC 3599 X--ray spectral fits.}
\begin{center}
\begin{tabular}{ccc}
\hline
\hline
   & diskbb+pow & pcf(pow) \\ 
\hline
$N_H$ ($10^{20}$) cm$^{-2}$& $2.8^{+0.2}_{-0.4}$    & $4.4^{+1.0}_{-0.7}$\\
PL $\Gamma$                          & $2.2^{+0.2}_{-0.2}$      & $3.6^{+0.1}_{-0.1}$\\
Diskbb norm.                            &$1900^{+780}_{-470}$ &--\\
\hline
Dataset & $T$ (eV) & Cov. fraction \\
\hline
ROS1 &  $83^{+4}_{-12}$& $0.14^{+0.07}_{-0.07}$\\
ROS2 &  $75^{+4}_{-4}$  & $0.69^{+0.30}_{-0.02}$\\
ROS3 &  $77^{+4}_{-4}$  & $0.95^{+0.01}_{-0.01}$\\
Cha1  & $71^{+3}_{-3}$   & $0.80^{+0.01}_{-0.01}$\\
Swi1   & $142^{+6}_{-7}$ & $<0.10$\\
Swi2   & $116^{+3}_{-6}$ & $<0.17$\\
Swi3   & $57^{+6}_{-16}$ & $0.98^{+0.01}_{-0.02}$\\
\hline
$\chi^2$ (dof) & 315.9 (303) & 627.4 (304) \\
nhp      & $0.29$                 & $10^{-24}$      \\
\hline
\hline
\end{tabular}
\end{center}
\medskip Errors were determined with $\Delta C=1$.
\label{xfit}
\end{table}

\begin{figure*}
\centerline{
\includegraphics[width=1.\textwidth]{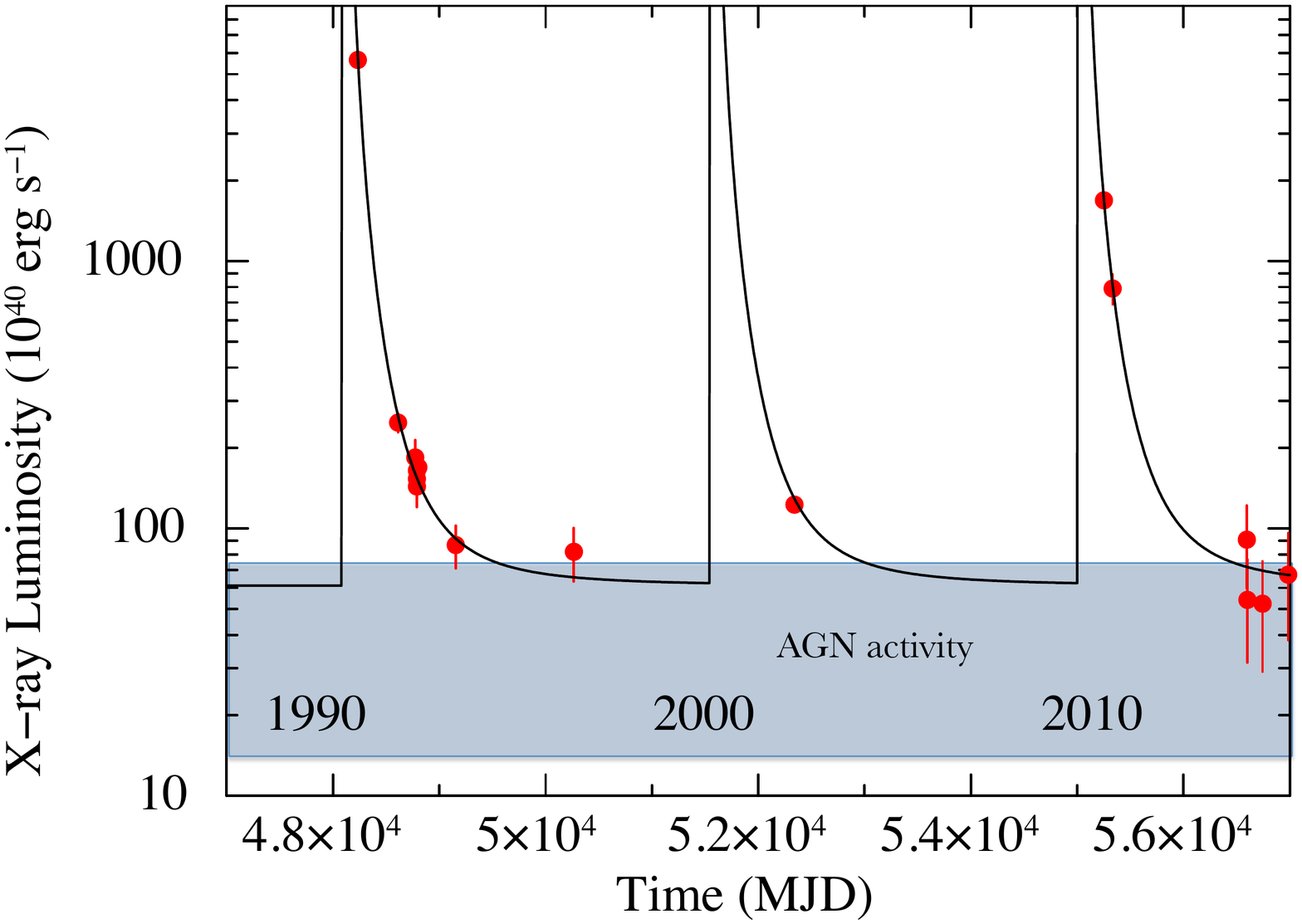}
}
\caption{
Long term X--ray luminosity light curve of IC 3599. Error bars are at $1\,\sigma$ confidence level.
{\it ROSAT}, {\it Chandra} and {\it Swift} count rates were converted into 0.01--10 keV unabsorbed luminosities by means of spectral fits 
assuming a source distance of 92 Mpc. 
The overall X--ray light curve has been fit with a $(t-t_0)^{-5/3}$ function repeating over a $P_0$ time. Free parameters are the starting date, 
the power law normalisation, the repetition time and a constant,  setting the basic emission level of IC 3599. 
The quiescent level (AGN activity) has been evaluated worsening the fit until a $10\%$ null hypothesis probability is attained.
This is nicely consistent with the mean value and the observed variability in the power law component used in the X--ray spectra fits.}
\label{fig2}
\end{figure*}

\section{Light curve fitting}

We fit together the luminosity light curve derived from all {\it Swift} X--ray data and archival {\it ROSAT} and {\it Chandra} data. 
To do this we computed the conversion factors from count rates to unabsorbed fluxes in the 0.01--10 keV energy band for each spectrum
based on the absorbed disc plus power law spectral model. These fluxes were then converted into luminosities adopting a distance of 92 Mpc 
for IC 3599 and used in Fig. 2.
Spectra that were obtained as the sum of different observations were then split into single observations and the same
conversion factor was applied to all of them. The flux of last {\it ROSAT} PSPC observation was estimated based on the spectrum of the 
third {\it ROSAT} PSPC observation (i.e. the closest in time). 
The flux of the RASS point was extrapolated to the 0.01--10 keV energy band based on the best fit spectrum (Grupe et al. 1995).
For the {\it ROSAT} HRI count rate we adopted a different approach. We fitted the last {\it ROSAT} PSPC spectrum with absorbed
single component models (power law, black body and bremsstrahlung). Based on these models we used PIMMS (v. 4.7b) to
extrapolate the {\it ROSAT} HRI rates to equivalent {\it ROSAT} PSPC rates. An average rate was computed and 
the standard deviation among the different models was added in quadrature. Then we converted these rates into fluxes as above.

The overall {\it Swift} light curve shows many similarities with the previous {\it ROSAT} light curve (Brandt et al. 1995; Grupe et al. 1995;
Komossa \& Bade 1999).
Both bolometric light curves can be fitted with a $t^{-5/3}$ power law, pointing to (at least) two different tidal disruption events.
The similarities in the decay of the events suggest that we are observing a recurrent phenomenon rather than 
the random occurrence of different accretion episodes.
The non-disruptive passage of the same star provides a more comprehensive explanation. 
%Two events separated by $\sim 20$ yr are clearly apparent in the data. 

We started by fitting the luminosity light curve with a different number of outbursts.
Values of the $\chi^2$ are reported in Table 3. It is readily apparent that the fit with 3 outbursts and a power law shape 
is superior to all the others. The improvement of a free power law index with respect to the fixed value $-5/3$ has been 
evaluated by means of an F-test. The probability of a random occurrence is $12\%$.

\begin{table*}[!tb]
\caption{X--ray luminosity light curve fitting.}
{\small
\begin{center}
\begin{tabular}{c|cc|ccc|ccc}
\hline
\hline     
                    & $t^{-5/3}$       & $t^{-5/3}$            & $t^{-\alpha}$  & $t^{-\alpha}$     & $\alpha$       & Exp.                  &  Exp.                    & $e-$ fold \\
                    & $\chi^2$ (dof)& nhp                      & $\chi^2$ (dof)& nhp                      &                        & $\chi^2$ (dof) &  nhp                     & (d) \\
\hline
2 outburst  & 37.3 (12)        &$2\times10^{-4}$& 25.3 (11)        &$8\times10^{-3}$&$-3.8\pm1.2$& 33.0 (11)         &$5\times10^{-4}$&$108\pm9$\\ 
3 outburst  & 9.3 (12)          &$7\times10^{-1}$& 7.1 (11)           &$8\times10^{-1}$&$-2.7\pm1.1$& 28.4 (11)         &$3\times10^{-3}$&$112\pm6$\\ 
4 outburst  & 29.0 (12)        &$4\times10^{-3}$& 24.3 (11)        &$1\times10^{-2}$&$-3.6\pm1.6$& 33.0 (11)         &$5\times10^{-4}$&$108\pm6$\\
\hline
\hline
\end{tabular}
\end{center}
}
\medskip
\label{lcurve}
\end{table*}
 
In addition to the luminosity light curve, we folded the temperatures derived from spectral analysis  along with the suggested 
orbital period. Disc temperature should evolve as $T\propto t^{-5/12}$, being $\mdot\propto t^{-5/3}$ and $T\propto \mdot^{1/4}$ 
(from accretion disc theory, Lodato \& Rossi 2010). Fitting the temperature evolution for three outburst peaks we obtain a reduced $\chi^2_{\rm red}=0.6$ 
for 6 degrees of freedom and a null hypothesis probability of $75\%$ (upper panel in Fig. 3). 
If we fit instead the temperature evolution for two outbursts only with the same model we derive a reduced $\chi^2_{\rm red}=32.0$ for 
6 degrees of freedom and a null hypothesis probability of $10^{-38}$ (orange dashed line in Fig. 3). If we add a constant to the power law 
model, mimicking the presence of an underlying quiescent accretion disc  we obtain $\chi^2_{\rm red}=6.6$ for 5 degrees of freedom 
and a null hypothesis probability of $4\times 10^{-6}$ (continuous line in the lower panel of Fig. 3). The constant quiescent disc 
temperature is $42\pm6$ eV. Note that such a constant temperature baseline might be present also in the first fit but it is not required by the data.
We note that the presence of three peaks depends on the {\it Chandra} data only. If these data are strongly affected timing and spectral 
variability unrelated to the TDE, a longer period (twice) should be envisaged.

Based on light curve and disc temperatures evolution we found strong indication for multiple (equal) tidal stripping events.
In particular, the timing and spectral data points toward three (rather than two) 
outbursts taking place during the 1990--2014 time span, with a recurrence time of 9.5 yr (see Figs. 2 and 3).
Assuming an accretion efficiency of $\eta=0.1$, from the outburst light curve we can estimate the peak mass accretion rate 
to be $\mdot_{\rm peak}\sim 0.01\msun$ yr$^{-1}$ and the accreted mass per episode to be $\Delta M\sim 2.5\times 10^{-3}\msun$.
Clearly these are lower limits, missing the early stages of all the outbursts. 
In addition, with these new ephemerides, radio observations took place only 2.6 yr after an outburst episode and can 
be accounted for by emission internal to the jet (van Velzen, K\"ording \& Falcke 2011).

\begin{figure}
\centerline{
\includegraphics[width=0.5\textwidth]{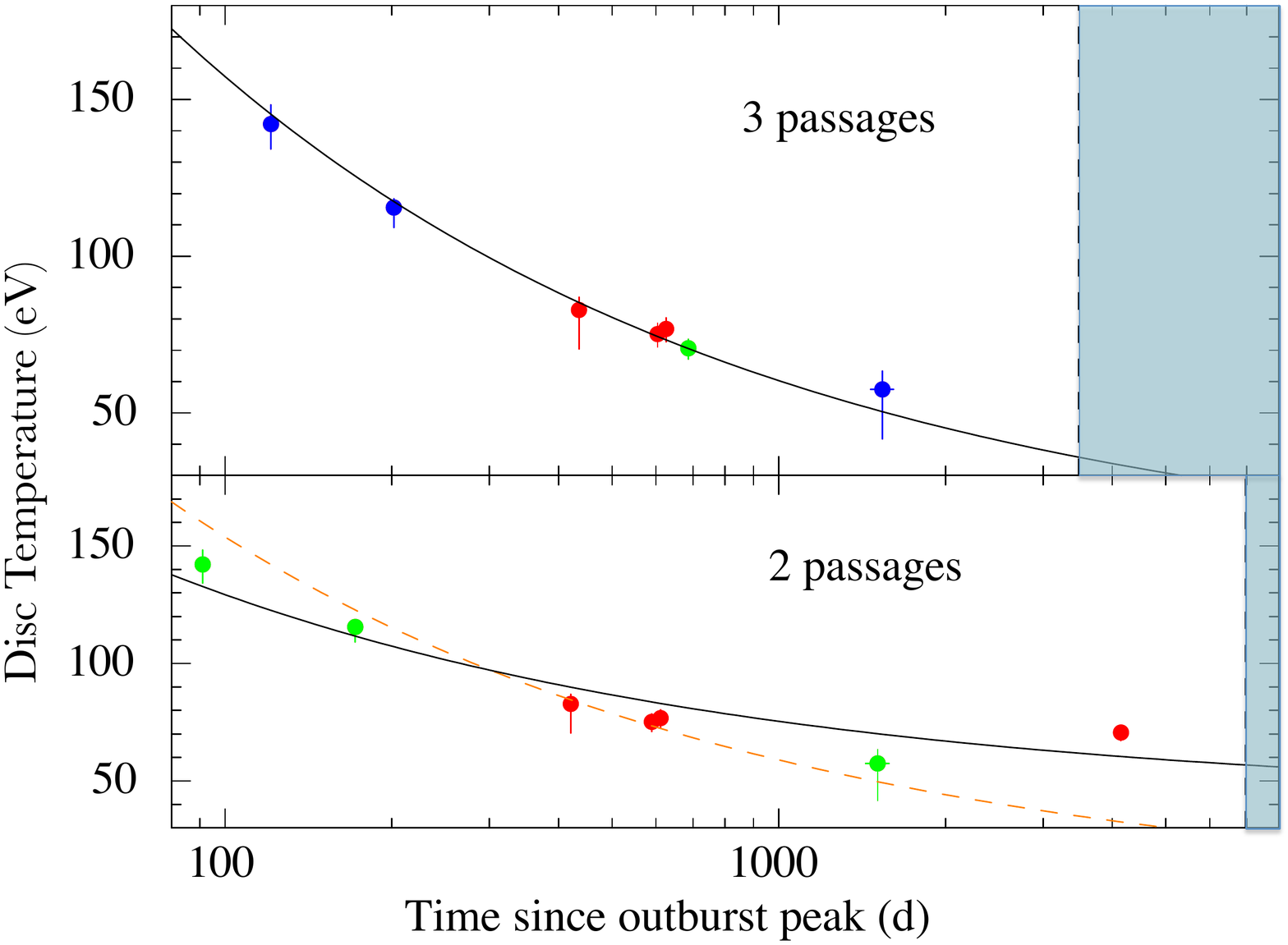}
}
\caption{ {\it Upper panel}: disc temperatures as derived from the spectral fits folded on a three passages light curve.
Error bars are at $1\,\sigma$ confidence level.
Red points ($ROSAT$) refer to the first passage, the green point ($Chandra$) to the second passage and blue points ($Swift$) to the 
third one. Disc temperature evolution is fitted with a fixed $T\propto t^{-5/12}$ power law (deriving from $\mdot\propto t^{-5/3}$ and 
$T\propto \mdot^{1/4}$ typical of an accretion disc and therefore being a direct test for the
disc cooling, Lodato \& Rossi 2010). 
{\it Lower panel}: as above but in the case of two passages. Red dots ($ROSAT$ and $Chandra$) refer to the first passage and 
green dots ($Swift$) to the second one. The fit with a $T\propto t^{-5/12}$ power law is shown as a dashed orange line. } 
\label{fig3}
\end{figure}

\section{Optical data}

\begin{figure}
\centerline{
\includegraphics[width=0.5\textwidth]{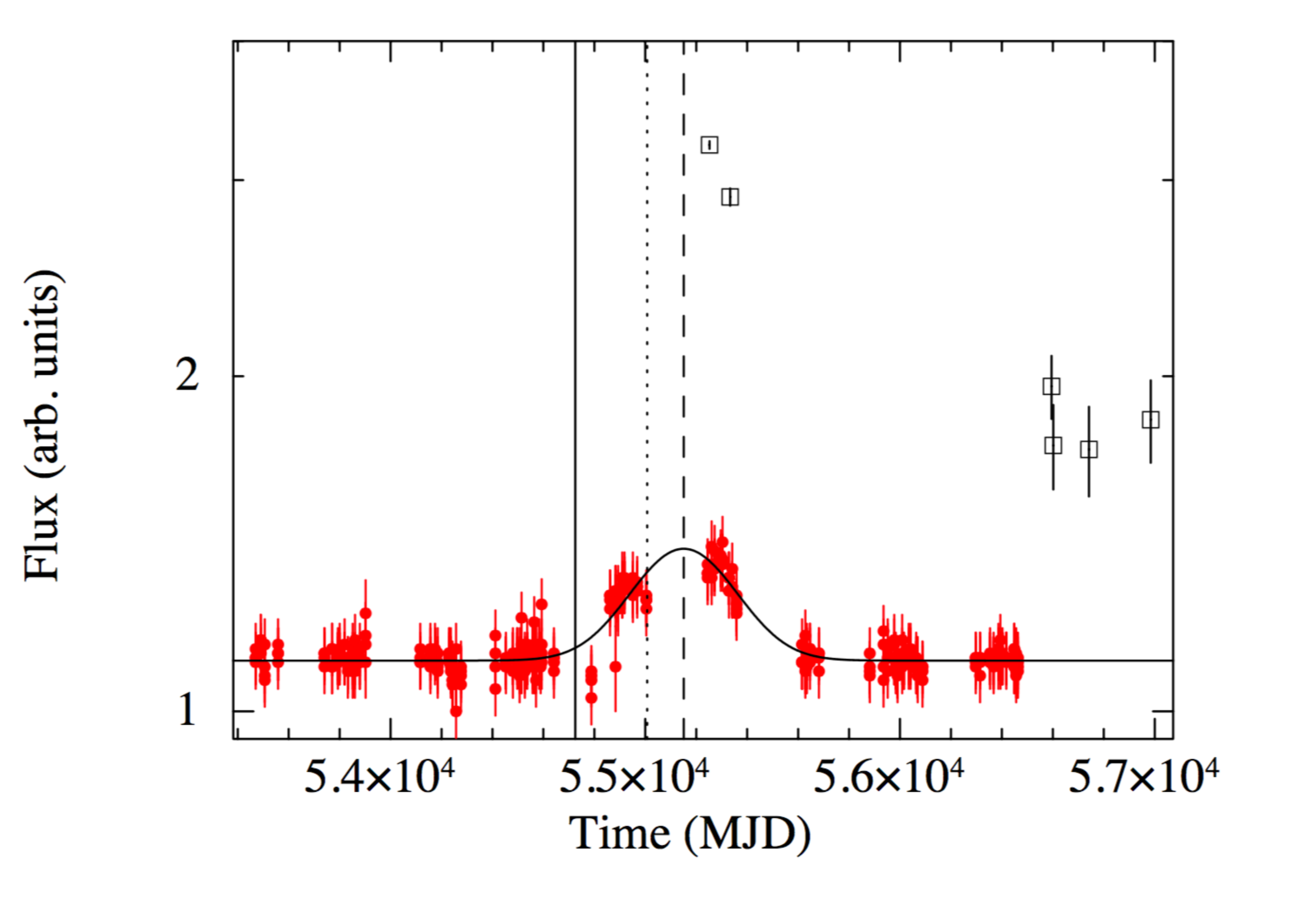}
}
\caption{Optical (flux) light curve of IC 3599 from the Catilina sky survey in arbitrary units (red dots), together with the {\it Swift}/XRT light curve (open squares) 
converted in flux using the spectral fit model, in arbitrary units. The Gaussian fit refers to the optical data. The vertical dashed line marks the time (MJD) of the optical peak, the dotted line the peak of the X--ray flux according to the multi-peak modelling. The continuous line marks the time at which the optical light curve 
rises by $10\%$ from the quiescent level.
} 
\label{cat}
\end{figure}

IC 3599 fell within the field of view of the Catalina sky survey and was monitored in the optical in the MJD 53470--56463 interval (Drake et al. 2009). 
The Catalina light curve shows a broad peak (see Fig. \ref{cat}). The light curve can be fit with a Gaussian centred on MJD $55151\pm11$ ($1\,\sigma$)
and with width of $200\pm18$ d. The peak of the optical emission occurs after the estimated peak of the X--ray emission by $\sim 
140$ d. However, due to the large width of the optical flare, the optical emission starts $\sim 280$ d before the high energy peak (assuming as the 
start of the optical flare when the optical flux rises by $10\%$ over the constant value).

This gives us the possibility to study in detail the outburst start (and prepare a follow-up strategy for the next outburst). 
The rise time in a TDE is governed by the circularisation time (i.e. the time to form the accretion disc) and the viscous time (i.e. the time
needed to transfer matter from the outer disc edge to the central compact source). Several papers have appeared recently on this subject
(Bonnerot et al. 2015; Guillochon \& Ramirez-Ruiz 2015; Piran et al. 2015).
Irrespective of which is the dominant mechanism, we can safely assume that the viscous time of the disc, $t_{\nu}$,  is shorter 
than the observed rise time. Given the optical light curve the time it takes to rise from $10\%$ of the quiescent flux to the peak flux is $\sim 420$ d.
Fitting the optical light curve with a symmetric exponential function, we derive an $e-$folding rise/decay time of $178\pm14$ d.
Based on Guillochon \& Ramirez Ruiz (2013), the viscous time can be expressed as:
$$
t_{\nu}=12\,\beta^{-3}\,M_*^{-1/2}\,R_*^{3/2}\,\alpha_{0.1}^{-1}\ {\rm d}
$$

\section{Constraining the orbit and star characteristics}
Partial disruption of stars has been investigated by means of hydrodynamical simulations (Guillochon \& Ramirez Ruiz 2013; 
Macleod et al. 2013; Macleod et al. 2012). 
Depending on the star structure (modelled as a polytrope with index $\gamma=4/3$ or 5/3), 
it has been shown that mass can be extracted from the star for impact parameter $\beta$ as low as 0.5--0.6 whereas complete destruction 
occurs for $\beta>0.90$ for $\gamma=5/3$ and for $\beta>1.85$ for $\gamma=4/3$, respectively (Guillochon \& Ramirez Ruiz 2013). 
In order to constrain the encounter and star characteristics we compare the mass accretion at peak rate $\mdot_{\rm peak}$ and the total amount
of matter $2\,\Delta M$ lost by the star during each passage with the corresponding quantities estimated through hydrodynamical 
simulations (Guillochon \& Ramirez Ruiz 2013). These quantities depend strongly on the impact parameter $\beta$ and, to a lesser extent, 
on the star mass and radius.

The peak mass accretion rate 
has been estimated based on the RASS data, converting the peak luminosity into a peak mass accretion rate  by assuming
a $10\%$ conversion efficiency. In order to bracket uncertainties the allowed parameter space 
is computed accepting values of the peak accretion rate in the interval 1--3 of observed value. 
The other parameter considered is the total amount of mass accreted during one
flare episode. This has been estimated by integrating the luminosity light curve over one orbital period. 
Also for this parameter we searched solutions in the interval 1--3 of the observed total mass accreted.
Finally, we require that the viscous time of the disc, $t_{\nu}$, is shorter than $178\pm42$ d ($3\,\sigma$).
A mass range of $0.2-100\msole$ and a (unrelated) radius range of $0.01-10^4\rsun$ were blindly searched. 
We consider two different star models based on a polytropic index of $\gamma=4/3$ (over the interval $\beta=0.5-4.0$) and 
$\gamma=5/3$ ($\beta=0.45-2.5$). Results are shown in Fig. 4 for the two indexes separately.

In case of $\gamma=4/3$ we find solutions in the range $\beta=0.58-0.72$. 
The range of allowed values of $\beta$ has been obtained without requiring a priori a partial disruption event, searching for solution 
in the $\beta=0.5-4.0$ interval. This results therefore strongly (and independently) supports that the flaring events in 
IC 3599 are related to a partial tidal disruption ($\beta<\beta_c=1.85$) of an orbiting star. 
The corresponding mass range is for $M_*>4\msole$. We blindly explored the mass-radius 
plane, but stars do not fill this plane homogeneously. To test the consistency of our findings with stellar models, we investigated the mass-radius relation 
by means of the Single Stellar Evolution (SSE) code (Hurley, Pols \& Tout 2000), assuming solar metallicity (we verified that a change 
in the metallicity does not change our results sensibly). With this additional constraint we have that the range of allowed masses 
reduces to $M_*=15-45\,\msole$ for stars on the main sequence and to $M_*=4-15\,\msole$ for slightly evolved stars.
The corresponding radii in the selected region are  $R_*\sim 5-9\,\rsun$ for main sequence stars and $R_*\sim 4-7\,\rsun$ for slightly evolved stars
(see Fig. 4 upper panel). 
The allowed eccentricities range in $e\sim 0.995-0.997$ and the pericentre distance $r_{\rm p}\sim 84-93\,r_{\rm g}$
(where $r_{\rm g}=G\,M_{\bullet}^{\rm IC}/c^2$ is the gravitational radius of the 
black hole in IC 3599, $G$ the gravitational constant and $c$ the speed of light).
The eccentricity is high but not unlikely given the predicted distribution of eccentricities around the central black hole in our 
Galaxy (Gillessen et al. 2009). This eccentricity is larger than the critical eccentricity for bound orbits below which all stellar debris 
remains bound and feed the black hole on a longer time scale (Hayasaki et al. 2013).

In the case of a polytropic star with index $\gamma=5/3$, we searched solutions for $\beta$ in the 0.45--2.5 range (see Fig. 4 lower panel). 
We do find solutions in a narrow range of $\beta= 0.49-0.55$. Again the allowed range of $\beta$ is well below the critical value for a complete 
tidal disruption ($\beta_c=0.9$). The allowed mass range is $M_*\gsim 1.5\msole$. However polytropic stars with $\gamma=5/3$, can satisfactorily 
describe Sun-like stars (or smaller) and we do not find therefore acceptable solutions in this case.

\begin{figure}
%\centerline{
\includegraphics[width=0.5\textwidth]{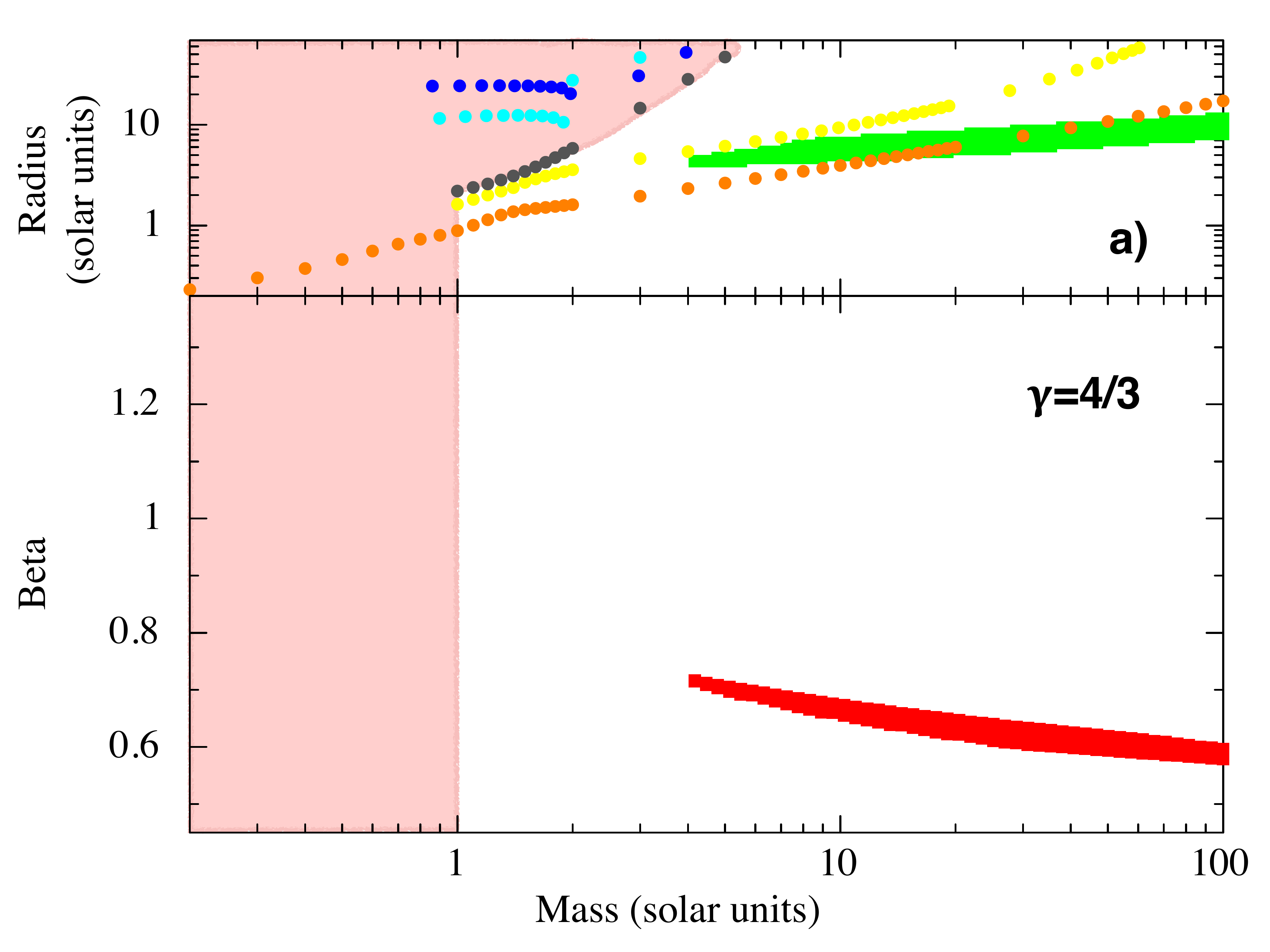}\\
\includegraphics[width=0.5\textwidth]{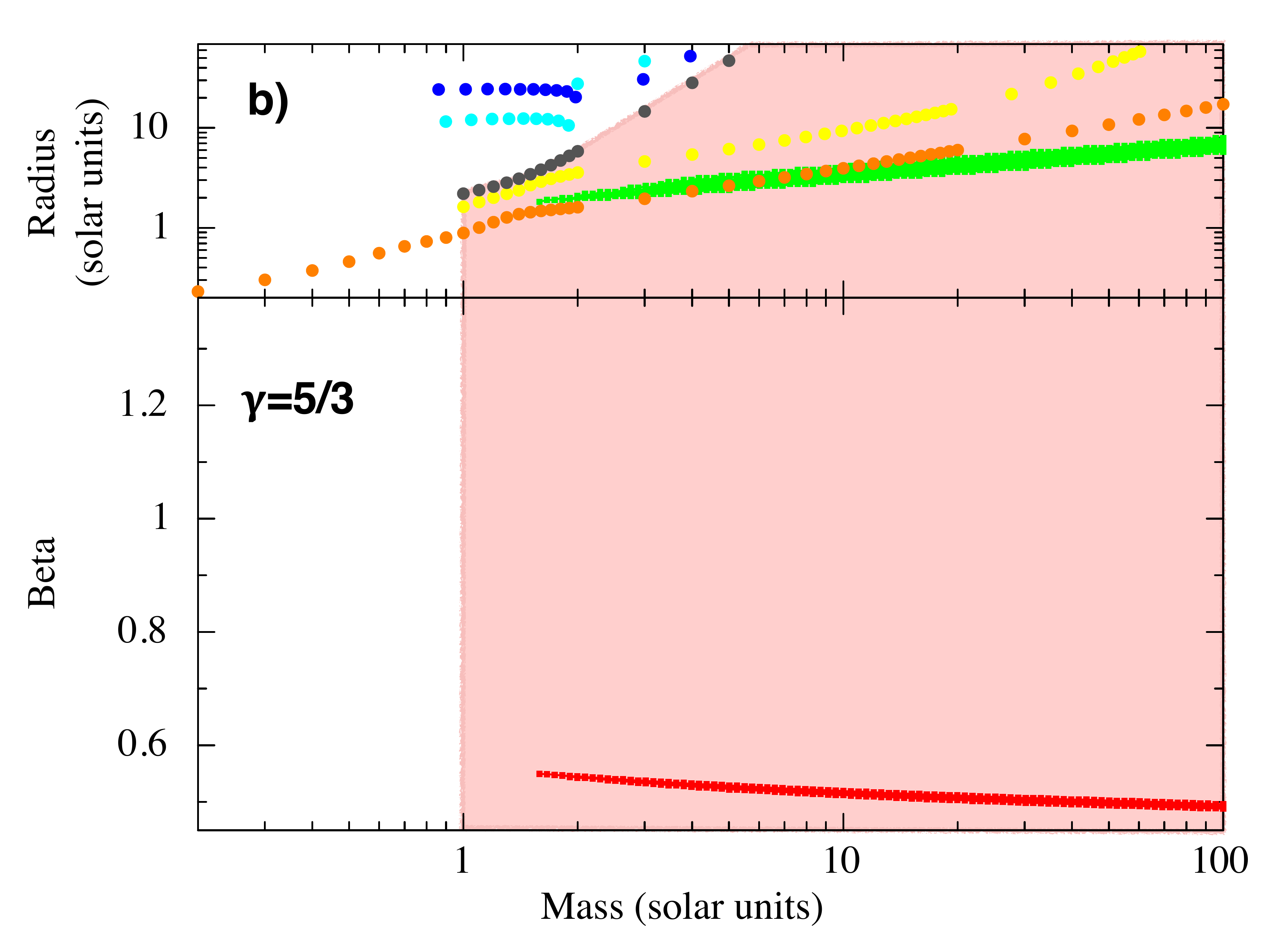}
%}
\caption{Radius of the orbiting star and impact parameter $\beta=r_{\rm t}/r_{\rm p}$ as a function of the star mass.
The allowed regions were computed based on fitting formulae in Guillochon \& Ramirez-Ruiz (2013 and related errata corrige). 
{\it Upper panel}: orbiting stars characterised by a polytropic index $\gamma=4/3$. The red  region in the lower panel shows the 
allowed $\beta$ parameters. The light red region approximately exclude the low mass end ($\lsim 1\msun$) and the stars more evolved 
than the giant branch because they are not well described by a polytropic index $\gamma=4/3$. Orange dots indicate stars in the main 
sequence phase, yellow dots the Hertzsprung gap, black dots the giant branch, light blue dots the helium core-burning, and blue dots the 
asymptotic giant branch phase. 
{\it Lower panel}: shows the case of a polytropic star with index $\gamma=5/3$. Colours, dots, and regions are as above. 
The light red region approximately excludes the high mass end ($\gsim 1\msun$), because these stars are not well described by a 
polytropic index $\gamma=5/3$ (but includes evolved giant stars). No acceptable solutions remain for this case.}
\label{fig4}
\end{figure}

\section{Conclusions}

We report here on the first candidate of periodic, partial tidal disruption events. These events were observed as large ($>100$) flux increases 
in the 24-yr long X--ray light curve of the close, low-luminosity active nucleus IC 3599 (see Fig. 2).  
We discuss several possibilities to explain these large flares and likely conclude that they cannot be ascribed to AGN variability:
single flares are unlikely, at $\gsim 4.0\,\sigma$ level, to come from known, observed extremely variable AGNs, and the flaring instability region
suggested by some accretion disc models (e.g. Honma et al. 1991) lies above the mean accretion rate observed in IC 3599 by  a factor of $\gsim 15$.
By modelling the light curve and the disc temperature evolution, as derived by spectral fits, we find out that three flares are statistically 
preferable over two, resulting in a $\sim 9.5$ yr periodicity. This cannot be appreciated from the light curve where the second flare is 
largely missed due to sparse observations, but results from the disc temperature evolution: if the standard disc model can apply to the observations of IC 3599,
then the 3-outbursts model is preferable over the 2-outburst model at $\sim 3.8\,\sigma$. Based on the disc temperature evolution the same is true at 
$\sim 4.3\,\sigma$ confidence level.
A tidal disruption model applied to IC 3599 provides a good match to the data and, independently, solutions for partial disruption events only.

In addition, we note that the average quiescent luminosity of the AGN is just a fraction of average flare luminosity ($\sim 20\%$),
leaving open the possibility that the overall AGN activity of IC 3599 is entirely spoon-fed by the orbiting star (MacLeod et al. 2013). 
%Tidal debris have lower energies than the star and are dispersed in a range of highly eccentric elliptical orbits, inside the star's orbit.  
Given the relatively large mass stripped every passage, the phenomenon is short-lived and based on our estimates it can last
$\lsim 10^4$ yr. The next passage should occur in 2019, giving us the ability to plan a detailed monitoring campaign to explore 
the characteristics of the orbiting star. 
The very short orbital period gave us the possibility to observe more than one passage. Other tidal disruption events presently known 
might be similar to this case but just with a longer orbital period. 

\begin{acknowledgements}
We acknowledge useful discussions with G. Tagliaferri, G. Ghisellini and R. Salvaterra. We thank R. Campana for useful 
discussions about power spectral densities and for his public python simulation software. 
\end{acknowledgements}

\appendix
\section{AGN variability}

Given the low-luminosity AGN nature of IC 3599 we investigated in details here if the observed flares can come from this activity.
We approached the problem either studying standard AGN variability and by taking an unbiased sample of the most variable AGNs 
as observed by the {\it Swift} satellite.

\subsection{Intrinsic AGN variability}

We simulated the long-term light curve of an AGN (Vaughan et al. 2003). We assumed a standard broken power law for the power 
spectral density (PSD) of a long term (years) monitoring light curve. We considered an index --1 at short frequencies and 
--2 at high frequencies, with a break at $10^3$ Hz scaled to the mass and luminosity of the black hole in IC 3599  (McHardy et al. 2006). 
We checked that even heavily changing these numbers the final result does not depend on them. We run a simulation 
generating a number of light curves based on the PSD above, including no root-mean-square (rms) variability and no background. 
The long-term light curve changes in all cases are less than a factor of a few with respect to the mean starting value (on timescales of days). 
We then turned on the rms variability and this value is the main driver for the variability. Even assuming a $100\%$ rms variability, 
we obtained a maximum increase by a factor of $\sim 6-7$ in the count rates (no background included) from the mean value on 
timescales longer than days. From this analysis we conclude that normal AGN variability is not able to produce the strong variability we observe.

\begin{figure}[!h]
\includegraphics[width=0.3\textwidth,angle=-90]{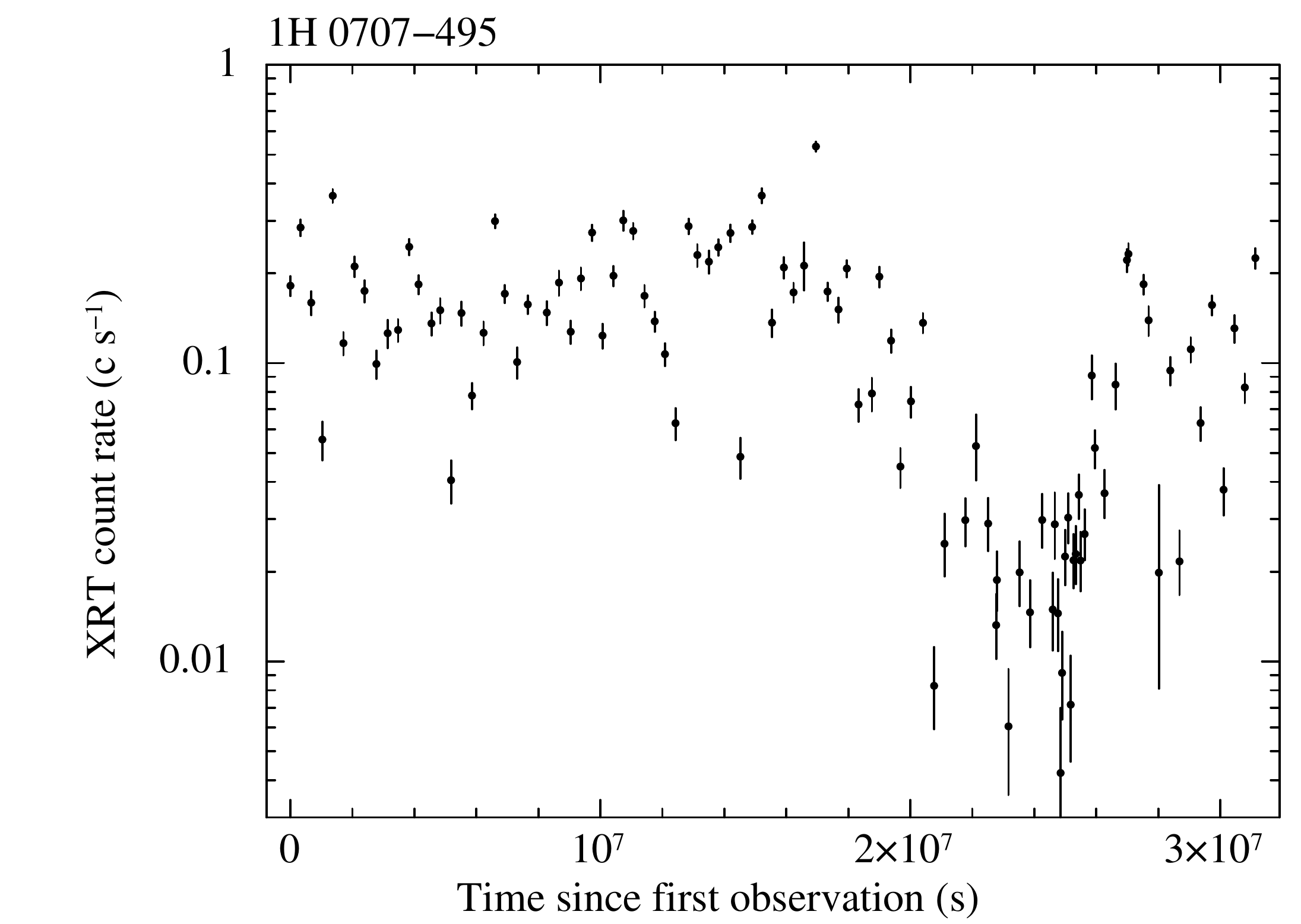} 
\includegraphics[width=0.3\textwidth,angle=-90]{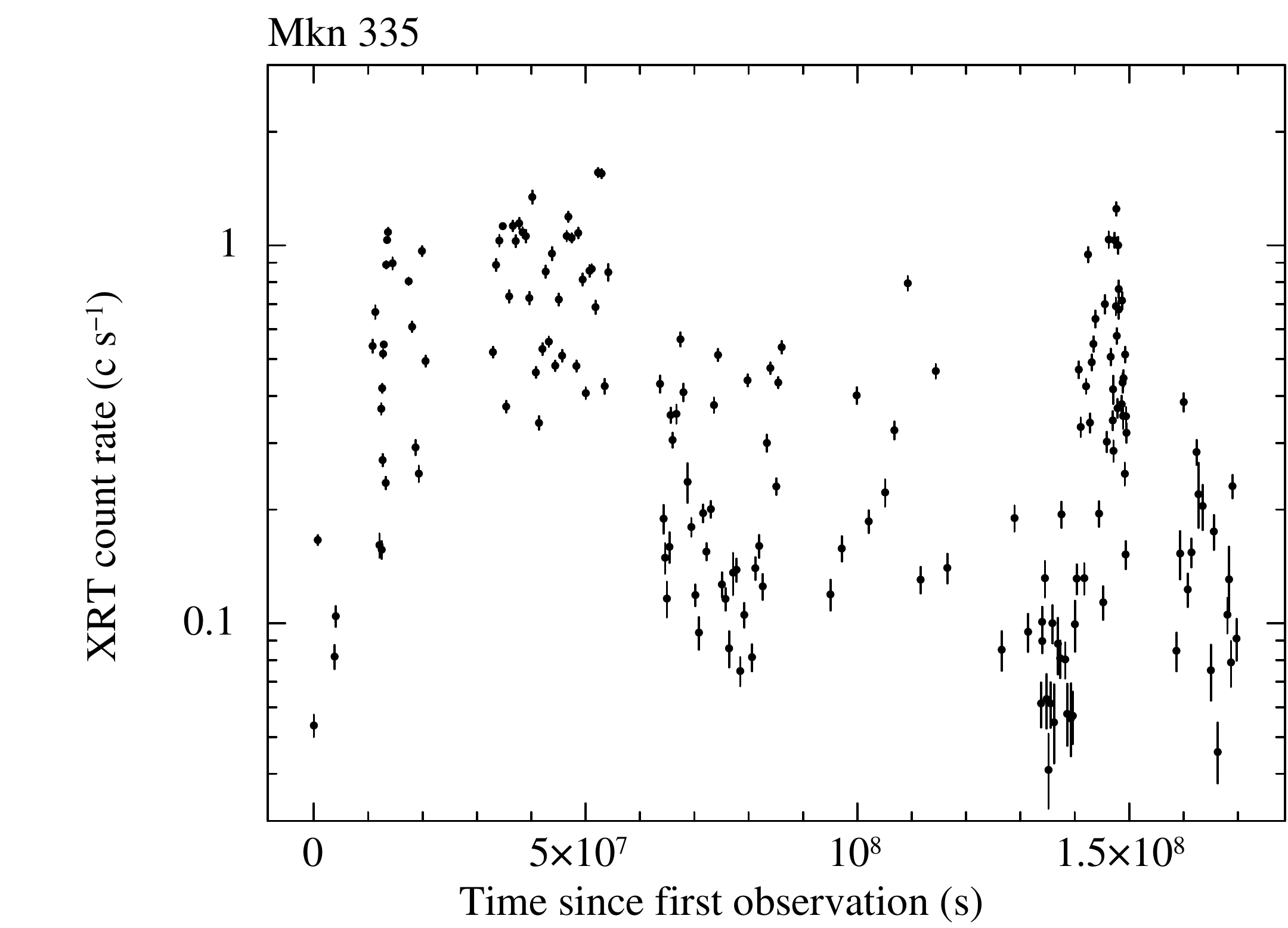} 
\includegraphics[width=0.3\textwidth,angle=-90]{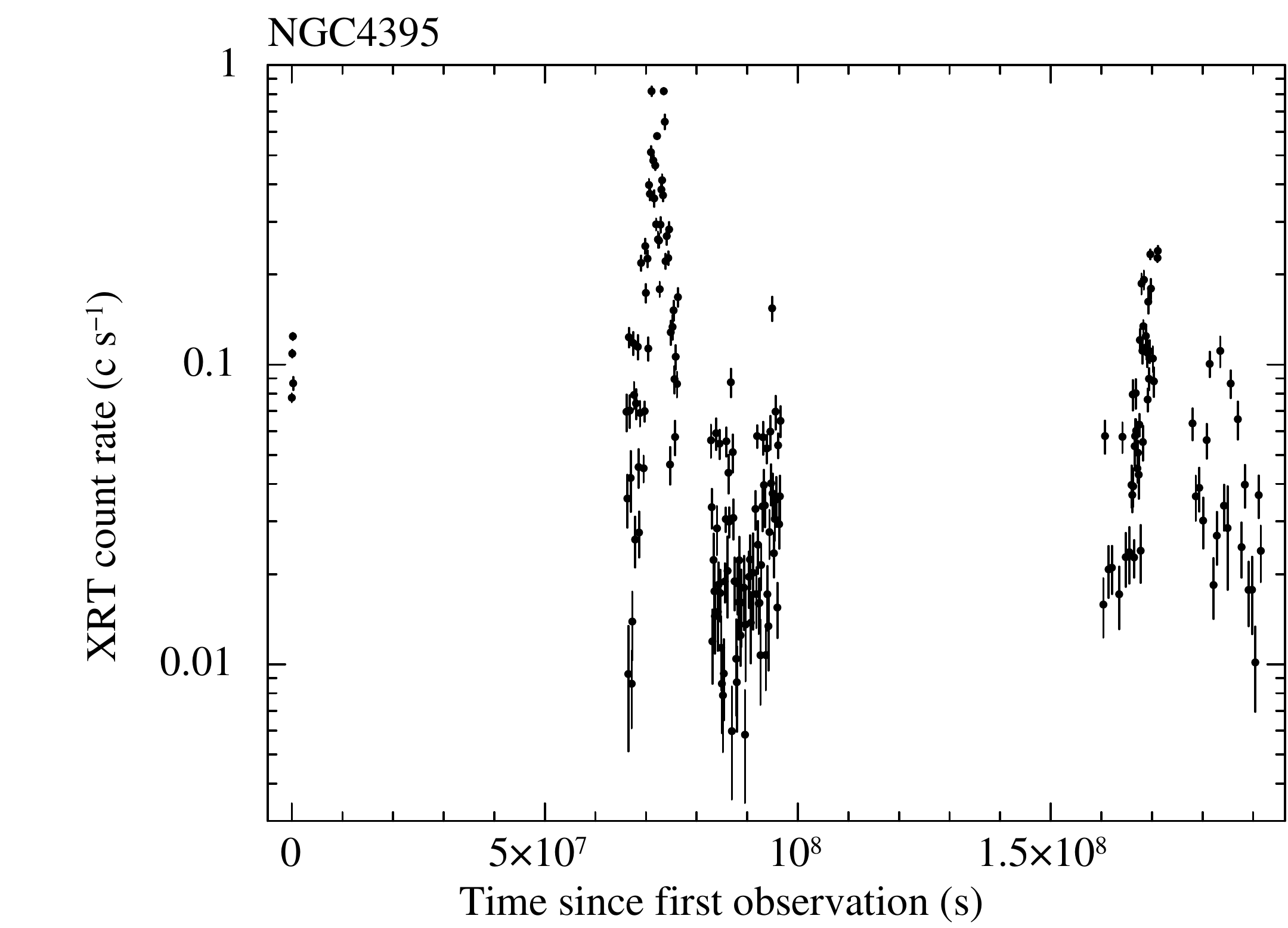} 
\includegraphics[width=0.3\textwidth,angle=-90]{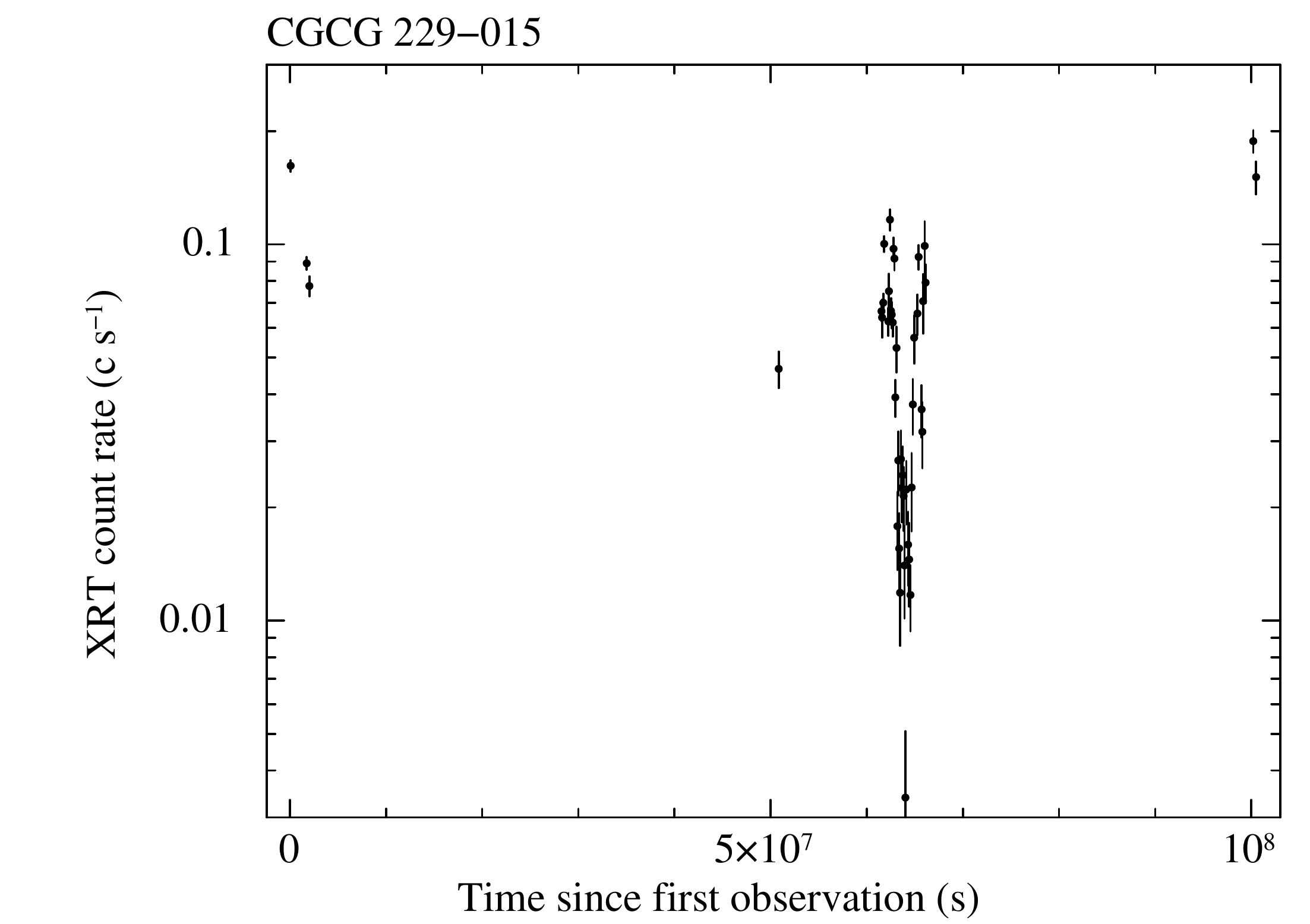} 
\caption{{\it Swift}/XRT light curves of the 4 selected highly variable AGN.}
\label{fig5}
\end{figure}

\subsection{AGN with large intrinsic variations}

We started from {\it Swift}/XRT light curves spanning a 10 yr basis. We first considered the AGN sample monitored by BAT, 
consisting of bright nearby AGNs with more than 5 {\it Swift}/XRT observations. In this sample there are just 3 AGNs showing a 
count rate change by more than a factor of 20. These are Mkn 335 ($\Delta = 42$), NGC 4395 ($\Delta = 141$) and CGCG 229--015 ($\Delta = 56$, 
even if in this case the change is a dip rather than a flare). One very efficient method to find highly variable AGN consists in comparing 
soft X--ray observations at different epochs. Grupe et al. (2001) showed that 4 out of 113 bright RASS objects observed also in ROSAT 
pointings showed dramatic variations. Grupe et al. (2012) added other 4 objects to this list by means of subsequent {\it XMM-Newton} and 
{\it Chandra} observations. All but two are Narrow Line Seyfert 1 galaxies. The first is RX J1624.9+7554, a non-active galaxy for 
which a TDE has been invoked (Grupe et al. 1999), the second one is IC 3599. In this Grupe-Komossa sample of highly variable AGNs there are 
then 8 objects but only 3 of them show variations larger than a factor of 30. These are 1H 0707--495 ($\Delta = 114$, dip-like variation), 
Mkn 335 (as above) and WPVS 007 ($\Delta = 36$). WPVS 007 is however too faint for Swift and the light curve is unusable because 
it has too few points (Grupe et al. 2013). We have then 4 objects with good-enough sampled light curves, showing flare-like features (Fig. A.1).

One conservative test is to extract randomly from these curves 7 points (the number of points comprising the first TDE flare) and 6 points 
(the number of points comprising the third TDE flare) separately, and fit them with a tidal decay 
template $(t-t_0)^{5/3}+c$ to mimic one TDE at a time. We simulated 500,000 light curve realisations for each source by randomly sampling the 
observed light curves, extracting 7 or 6 points, respectively.
We then excluded simulated light curves with a rate variation less than a factor of 30 (in order not to fit constants, note that even if the flux variation in the 
{\it Swift}/XRT data is $\sim 30$ the count rate variation is $\sim 100$ due to spectral variability, therefore our rate change provides a conservative 
estimate) and fit the remaining light curves with the above model. 
We adopted two different approaches. In the first one we counted how many simulations have a null hypothesis probability larger than $5\%$ and, 
based on this number, evaluated the probability of obtaining by chance a TDE-like light curve.
In the second approach we took among all these light curves the best one in terms of $\chi^2$ and derived the 
probability of this $\chi^2$ with respect to the number of degrees of freedom (5 and 4  in our case, respectively). We do this because even 
the best selected $\chi^2$ is, for some sources, not good, at variance with the fit of the IC 3599 flare. This $\chi^2$ probability is then weighted for the number 
of trials (500,000). With these numbers we assessed the probability of randomly extract a TDE-like event from these light curves with 
a count rate increase by a factor of $> 30$ (the {\it Swift} rate variation is $\sim 100$). We obtained in this way conservative limits that are shown in Table A.1.
%As a final test, we took the central part of the Mkn 335 flare, cut the light curve from the peak, to the end of the observing season 
%(800 d in the figure) and chose the peak as the first point of each simulated light curve. Then, we run again the Monte Carlo simulation. 
%In this highly unlike case (at least we have selected 6 points instead of 7), we were able to reproduce the first IC 3599-like flare 
%(not saying about the second) with a significance of $4.47\,\sigma$.

A different check can be made on the spectral properties of IC 3599. In Fig. A.2 we show the spectral evolution of the five sources in 
our sample. This is quantified in terms of hardness ratio (computed as $HR=(H-S)/(H+S)$, being $H$ and $S$ the counts in the 2--10 keV and
0.5--2 keV energy bands, respectively, so that $HR$ can vary among --1 and 1) versus cont rates normalised to the minimum observed rate.
In this plane IC 3599 stands alone, reaching complete softness ($HR=-1$) at maximum and showing a clear spectral evolution.
1H 0707--495 is similarly soft but it does not show a marked spectral variability as the count rate changes. Mkn 355 and NGC 4395 show
spectral marked changes but never appear as very soft sources.

We should safely conclude that the flares observed in IC 3599 do not likely come from known (observed) AGN variability.

\begin{table*}
\caption{Simulated light curves from Swift/XRT variable AGN searching for random TDE-like events.}
\begin{center}
\begin{tabular}{cccccc} 
\hline
Source                   & TDE1 probab. & TDE3 probab.   & Num. Obs.  & Variation       & Type \\
                                &($\sigma$)        &($\sigma$)          &                      &                        & \\
\hline
1H 0707--495      & $  4.11 (4.95)$  & $  4.06 (4.84)$  &101             &$\Delta=114$& dip\\
Mkn 335                & $>4.75 (7.93)$ & $>4.75 (7.89)$  &179             &$\Delta=42$  & flare\\
NGC 4395            & $>4.75 (5.74)$ & $  4.06 (4.94)$   &190             &$\Delta=141$& flare\\
CGCG 229--015 & $>4.75 (>8.31)$& $>4.75 (>8.31)$&44                &$\Delta=56$  &dip\\
\hline
\end{tabular}
\end{center}

Probability of the first and third IC 3599 flares to be obtained by sampling the {\it Swift}/XRT light curves of highly variable AGNs.
The first probability is obtained by requiring that the simulated light curve has a null hypothesis probability larger than $5\%$. The
second probability (in parenthesis) is obtained by weighing the best fit light curve probability with the number of simulations carried out.
\end{table*}

\begin{figure}[!ht]
\centerline{\includegraphics[width=0.35\textwidth,angle=-90]{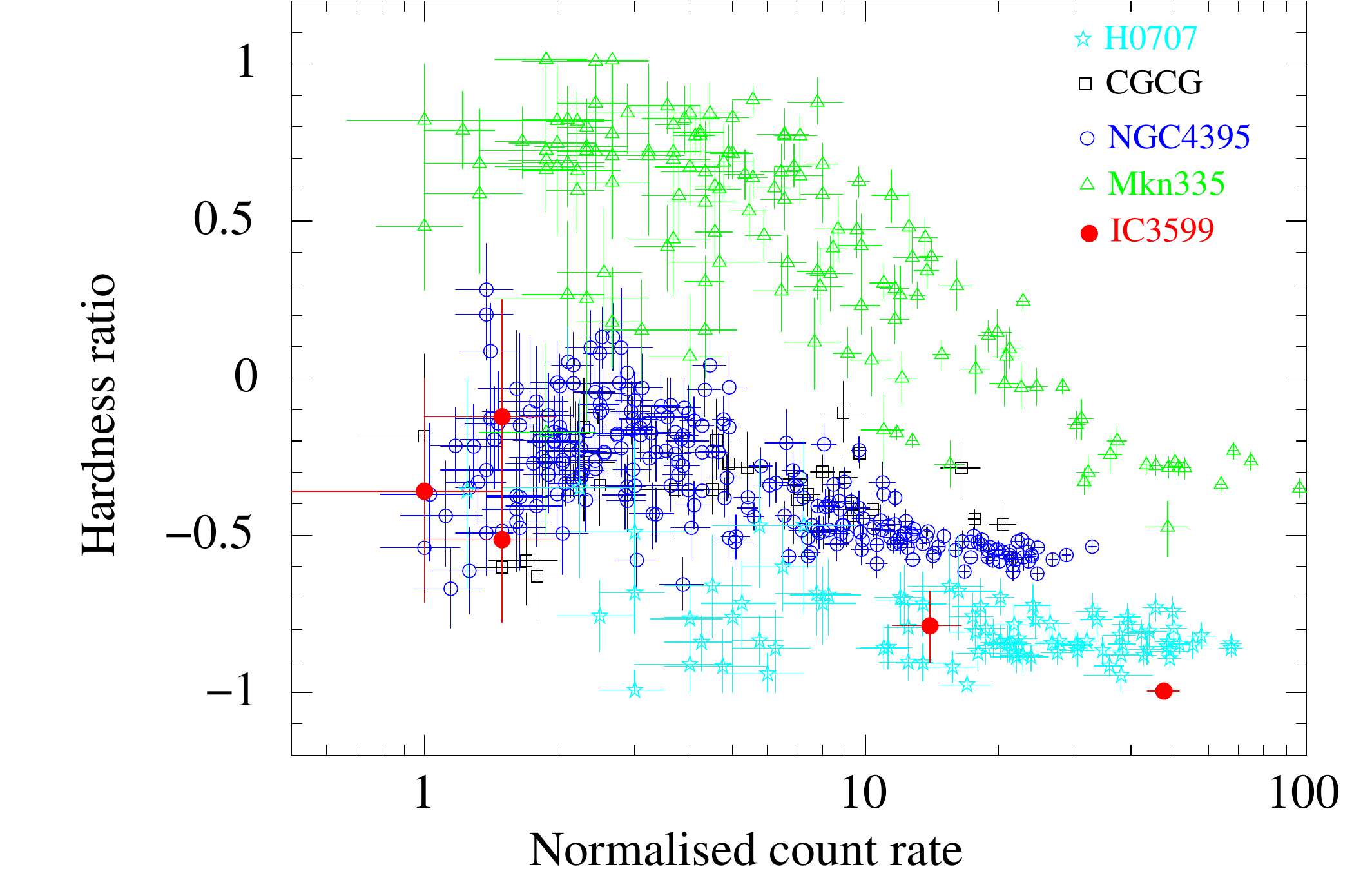}}
\caption{Hardness ratio as a function of the normalised count rates for the variable AGNs described above. 1H 0707--495 data are shown as 
light blue open stars, CGCG 229--015 as black open squares, NGC 4395 as blue open circles, Mkn 335 as green open triangles, and IC 3599 
as filled red circles. Count rates are normalised to the minimum rate observed. The hardness ratio is computed as the ratio as ($H-S)/(H+S)$
being $H$ the number of counts in the 2--10 keV energy band and $S$ those in the 0.5--2 keV band.}
\label{fig6}
\end{figure}

\subsection{Disk instability models}

The standard geometrically thin, optically thick accretion disc model has thermally unstable regions if the accretion rate is larger than a 
critical value. A new branch of equilibrium solutions should exist for rate across the Eddington luminosity, named slim discs (Abramowicz et al. 1988).
Some works have shown that a limit cycle might take place, resulting in a flare-like behaviour (Honma et al. 1991). Magneto-Hydrodynamical 
simulations showed, on the contrary, that such disks could be thermally stable (Hirose, Krolik \& Blaes 2009).
Observationally, Galactic transient X--ray binaries in outburst do not show any evidence for such an instability, except perhaps 
GRS 1915+105 (Xue et al. 2011). 

The mean mass accretion rate of IC 3599 is $\sim 0.005$ in Eddington units (for an efficiency of $10\%$,
where the peak accretion rate reaches $\sim 0.1$ the Eddington rate), far away from border of the 
unstable region, starting at a critical mass accretion rate $>0.1$ in Eddington units (Honma et al. 1991). 
Flare models do not make a clear prediction of the shape of the flare but make key prediction
that the flare width is a very short fraction of the times between two different flares (Xue et al. 2011). The ratio of the FWHM of the flare 
with respect to the duration time between two consecutive flares must be $\lsim 3.6\%$.
We modelled the two main peaks of IC 5399 with a Gaussian, limiting the maximum luminosity to a few times the Eddington value, 
and found a FWHM$\sim 500$ d (leaving free the maximum luminosity provides a better fit and an even larger FWHM$\sim 1200$ d). 
Taking the times among the two main flares, we obtain a ratio of  $\sim 7\%$ ($\sim 17\%$ for a free Gaussian shape).
This is more than double what expected based on models.

We also note that two possible very recent flare-like events in AGNs, XMM SL1 J061927.1--655311 (Saxton et al. 2014) and 
NGC 2617 (Shappee et al. 2014), involve smaller flux variations ($\lsim 20$) and much more complicated X--ray light curves.

\section{UVOT analysis}

\begin{figure*}[!ht]
\centerline{
\includegraphics[width=0.8\textwidth]{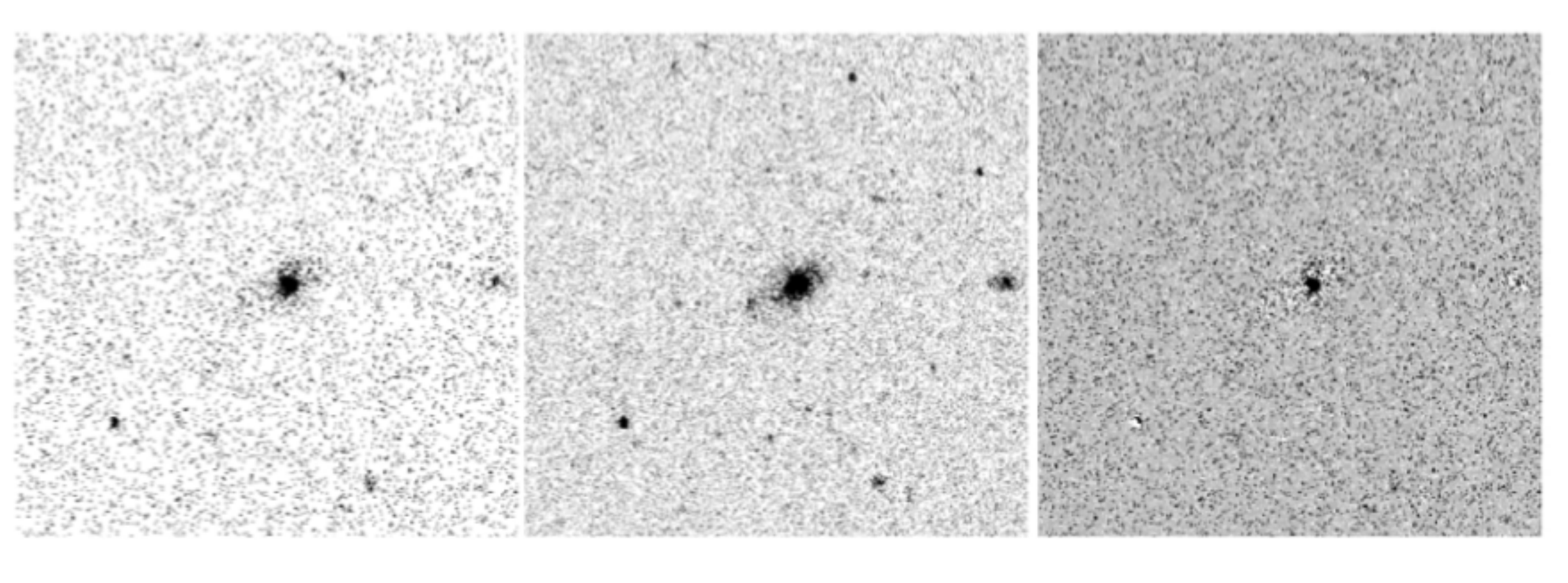}}
\caption{{Swift/UVOT images of IC 3599, obtained with the uvm2 filter on 2010
Feb 25 (left panel) and 2014 March 26 (central panel). The images are $5'\times 5'$, North is up, East is left. 
The right panel shows the result of digital image subtraction between these two images.
A clear residual is visible at a position consistent with the IC 3599 galaxy centre.}}
\label{fig7}
\end{figure*}

\subsection{Image analysis}

Together with XRT exposure, the {\it Swift} satellite took contemporaneous images of IC 3599 with the UVOT instrument.
During all the six {\it Swift} observations (but the fourth), UVOT observed IC 3599 with all its 6 filters (in the fourth only with the uvm2 filter).
Sky images were considered and fluxes were obtained with the {\tt uvotsource} task using the latest calibrations (Poole et al. 2008; Breeveld et al. 2011).
For UV images a circular extraction region of $8$ arcsec radius has been considered and for optical image a $5$ arcsec radius, respectively.
A background region close to IC 3599 free of sources and of $20$ arcsec radius has been selected.
Results are shown in Table B.1. A clear decrease in flux is apparent, especially at UV wavelengths with a decrease by a factor of 
$\sim 2$ (uvm2).

\begin{table}[!htb]
\caption{IC 3599 UVOT data analysis.}
\begin{center}
\begin{tabular}{ccc}
\hline
\hline
Time   & Flux \\
(MJD) & $10^{-15}$ erg cm$^{-2}$ s$^{-1}$ \AA$^{-1}$\\
\hline
v band & (5468 \AA)\\
55252.3  &$   2.23 \pm 0.11$\\
55333.0  &$   2.00 \pm 0.09$\\
56594.9  &$   1.70 \pm 0.10$\\
56741.9  &$   1.61 \pm 0.08$\\
\hline
b band & (4392 \AA)\\
55252.3  &$   2.20  \pm0.09$\\ 
55333.0  &$   2.03  \pm0.07$\\
56594.9  &$   1.63  \pm0.08$\\
56741.9  &$   1.58  \pm0.06$\\
\hline
u band & (3465 \AA)\\
55252.3  &$   1.97  \pm0.08$\\
55333.0  &$   1.79  \pm0.07$\\
56594.9  &$   0.99  \pm0.06$\\
56741.9  &$   1.03  \pm0.05$\\
\hline
uvw1 band & (2600 \AA)\\
55252.3  &$   2.43  \pm 0.10$\\
55333.0  &$   2.20  \pm 0.08$\\
56594.9  &$   1.09  \pm 0.06$\\
56741.9  &$   1.19  \pm 0.06$\\
\hline
uvm2 band & (2246 \AA)\\
55252.3  &$   2.59 \pm 0.12$\\
55333.0  &$   2.55 \pm 0.11$\\
56594.9  &$   1.16 \pm 0.08$\\
56601.9  &$   1.22 \pm 0.04$\\
56741.9  &$   1.23 \pm 0.04$\\
\hline
uvw2 band & (1928 \AA)\\
55252.3  &$   2.90  \pm0.09$\\
55333.0  &$   2.71  \pm0.08$\\
56594.9  &$   1.43  \pm0.06$\\
56741.9  &$   1.33  \pm0.05$\\
\hline
\hline
\end{tabular}
\end{center}
\medskip Errors are at $1\,\sigma$ confidence level.
\label{ottico}
\end{table}

\subsection{Astrometry}

In order to determine precisely the position of the transient source observed in IC 3599, 
we carried out digital image subtraction between the first and the last UV frames (uvm2) obtained with 
the Swift/UVOT. 
In the first image, obtained on 2010 February 25, the transient source is at its 
maximum while in the last epoch, obtained on 2014 March 26, it has faded below the host galaxy 
level (Table B.1). Both images were obtained with the
uvm2 filter and have exposure times of 0.2 ks and 1.2 ks, respectively. 
Before subtraction, the two images were aligned. We accounted for the difference
in the exposure times by multiplying the image obtained on Feb 2010 by a factor 
of six. As can be seen in the third panel of Fig. B.1, the result of the
image subtraction is good, showing a clear residual at the position of IC 3599. The
position of this residual is RA, Dec (J2000): 12:37:41.18, +26:42:27.0 (with an
uncertainty of  $0.3$ arcsec, calibrated against the USNOB1.0 catalogue). At the distance of IC 3599 
$1$ arcsec corresponds to 0.4 kpc.
The residual is located at $0.8\pm 0.4$ arcsec ($1\,\sigma$ confidence level) with respect to the position 
of the IC 3599 galaxy centre and is fully consistent with the position of the 
candidate TDE radio counterpart (Bower et al. 2013). The consistency of the position of the UV transient 
with the IC 3599 galaxy centre (at the $2\,\sigma$ level) is in agreement with a TDE origin.


\begin{thebibliography}{100}

\bibitem[]{Abramowicz88}
Abramowicz, M. A., Czerny, B., Lasota, J. P., Szuszkiewicz, E. 1988, ApJ, 332, 646

\bibitem[]{Bloom11}
Bloom, J. S. , et al. 2011, Sci, 333, 203

\bibitem[]{Bonnerot15}
Bonnerot, C., et al. 2015, MNRAS submitted (arXiv :1501.04635)

\bibitem[]{Bower13}
Bower, C. B., Metzger, B. D., Cenko, S. B., Silverman, J. M.,  Bloom, J. S. 2013, AJ, 763, 84

\bibitem[]{Brandt95}
Brandt, W. N., Pounds, K. A., Fink, H. 1995, MNRAS, 273, L47

\bibitem[]{Breeveld11}
Breeveld, A. A., Landsman, W.,  Holland, S. T.,  Roming, P. ,  Kuin, N. P. M.,  Page, M. J. 2011,  AIPC,  1358, 373 

\bibitem[]{Burrows11}
Burrows, D. N., et al. 2011, Nat, 476, 421

\bibitem[]{Cenko12}
Cenko, S. B., et al. 2012, ApJ, 753, 77

\bibitem[]{Churazov96}
Churazov, E., Gilfanov, M., Forman, W., Jones, C. 1996, ApJ 471, 673

\bibitem[]{Donley02}
Donley, J. L. , Brandt, W. N., Eracleous, E. Boller, Th. 2002, AJ, 124, 1308

\bibitem[]{Drake09}
Drake, A. J., et al. 2009, ApJ, 696, 870

\bibitem[]{Evans89}
Evans, C. R.,  Kochanek, C. S. 1989, ApJ, 346, L13

\bibitem[]{Gezari12a}
Gezari, S. 2012, EPJWC, 39, 3001

\bibitem[]{Gezari12}
Gezari, S., et al. 2012, Nat, 485, 217

\bibitem[]{Gillessen09}
Gillessen, S., Eisenhauer, F., Trippe, S., Alexander, T., Genzel, R., Martins, F., Ott, T. 2009, ApJ, 692, 1075

\bibitem[]{Gillessen12}
Gillessen, S., et al. 2012, Nat, 481, 51

\bibitem[]{Grupe12}
Grupe, D., Komossa, S.,  Leighly, K. M. , Gallo, L. C. 2012, EPJWC, 39, 6001

\bibitem[]{Grupe01}
Grupe, D., Thomas, H.-C., Beuermann, K. 2001, A\&A, 367, 470

\bibitem[]{Grupe99}
Grupe, D., Thomas, H.-C., Leighly, K. M. 1999, A\&A, 350, L31

\bibitem[]{Grupe95}
Grupe, D., et al. 1995, A\&A, 299, L5 

\bibitem[]{Grupe13}
Grupe, D., et al. 2013, AJ, 146, 78

\bibitem[]{Grupe15}
Grupe, D., Komossa, S., Saxton, R. 2015, ApJ, 803, L28

\bibitem[]{Guillochon14}
Guillochon, J., Loeb, A., MacLeod, M., Ramirez-Ruiz, E. 2014, ApJ, 786, L12

\bibitem[]{Guillochon13}
Guillochon, J. Ramirez-Ruiz, E. 2013, ApJ  767, 25

\bibitem[]{Guillochon15}
Guillochon, J. Ramirez-Ruiz, E. 2015, ApJ  submitted (arXiv:1501.05306)

\bibitem[]{Hayasaki13}
Hayasaki, K., Stone, N., Loeb, A. 2013, MNRAS 434, 909

\bibitem[]{Hirose09}
Hirose, S.,  Krolik, J. H., Blaes, O, 2009, ApJ, 691, 16

\bibitem[]{Ho08}
Ho, L. C. 2008, ARA\&, 46, 475

\bibitem[]{Honma91}
Honma, F., Matsumoto, R., Kato, S. 1999, PASJ,  43, 147

\bibitem[]{Hopkins06}
Hopkins, P. F., et al. 2006, ApJS, 163, 1

\bibitem[]{Hurley00}
Hurley, J. R., Pols, O. R., Tout, C. A. 2000, MNRAS, 315, 543

\bibitem[]{Ivanov01}
Ivanov, P. B.,  Novikov, I. D. 2001, ApJ, 549, 467
		
\bibitem[]{Kalberla05}
Kalberla, P. M. W. , et al. 2005, A\&A, 440, 775

\bibitem[]{Karas07}
Karas, V., Subr, L. 2007, A\&A, 470, 11

\bibitem[]{Komossa99}
Komossa, S., Bade, N. 1999, A\&A, 343, 775

\bibitem[]{Komossa12}
Komossa, S. 2012, EPJWC, 39, 2001

\bibitem[]{Lacy82}
Lacy, J. H., Townes, C. H.,  Hollenbach, D. J. 1982, ApJ, 262, 120 

\bibitem[]{Lasota11}
Lasota, J.-P., et al. 2011, ApJ, 735, 89 

\bibitem[]{Lodato09}
Lodato, G., King, A. R.,  Pringle, J. E. 2009, MNRAS, 392, 332

\bibitem[]{Lodato10}
Lodato, G., Rossi, E. M. 2010, MNRAS,  476, 359

\bibitem[]{Macleod13}
MacLeod, M., Ramirez-Ruiz, E., Grady, S., Guillochon, J. 2013, ApJ, 777, 133 

\bibitem[]{Macleod12}
MacLeod, M. Ramirez-Ruiz, E. Guillochon, J. 2012, ApJ, 757, 134

\bibitem[]{Marconi03}
Marconi, A., Hunt, L. 2003, ApJ, 589, L21

\bibitem[]{McHardy06}
McHardy, I. M., Koerding, E., Knigge, C. Uttley, P, Fender, R. P. 2006, Nat, 444, 730 

\bibitem[]{McHardy04}
McHardy, I. M., Papadakis, I. E., Uttley, P, Page, M. J.,  Mason, K. O. 2004, MNRAS, 348, 783

\bibitem[]{Mendel15}
Mendel, I., Levin, Y. 2015, ApJ, 805, L4

\bibitem[]{Nelson00}
Nelson, C. H. 2000, ApJ, 544, 91

\bibitem[]{Perets07}
Perets, H. B., Hopman, C., Alexander, T. 2007, ApJ, 656, 709	

\bibitem[]{Phinney89}
Phinney, E. S. 1989,  in Proc. 136th IAU Symp. 'The Center of the Galaxy'

\bibitem[]{Piran15}
Piran, T., et al. 2015, ApJ submitted (arXiv:1502.05792)

\bibitem[]{Poole08}
Poole, T. S., et al. 2008, MNRAS, 383, 627

\bibitem[]{Rees88}
Rees, M. J. 1988, Nat, 333, 523

\bibitem[]{Renzini95}
Renzini, A., et al. 1995, Nat, 378, 39

\bibitem[]{Sani10}
Sani, E., et al. 2010, MNRAS, 403, 1246 

\bibitem[]{Saxton14}
Saxton, R. D., et al. 2014, A\&A, 572, A1

\bibitem[]{Shappee14}
Shappee, B. J., et al. 2014, ApJ, 788, 48

\bibitem[]{Strubbe09}
Strubbe, L. E., Quataert, E. 2009, MNRAS, 400, 2070

%\bibitem[]{Strubbe11}
%Strubbe, L. E. , Quataert, E. 2011, MNRAS, 415, 168

\bibitem[]{Tanaka13}
Tanaka, T. L.  2013, MNRAS, 434, 2275

\bibitem[]{Ulmer99}
Ulmer, A. 1999, ApJ, 514, 180

\bibitem[]{vanVelzen11}
van Velzen, S.  K\"ording, E.  Falcke, H. 2011, MNRAS, 417, L51

\bibitem[]{Vaughan04}
Vaughan, S., Edelson, R., Warwick, R. S. 2004, MNRAS, 349, L1 

\bibitem[]{Vaughan03}
Vaughan, S., Edelson, R., Warwick, R. S., Uttley, P. 2003, MNRAS,  345, 1271

\bibitem[]{Xue11}
Xue, L., Sadowski, A., Abramowicz, M. A.,  Lu, J.-F. 2011, ApJS, 195, 7

%\bibitem[]{Perets09}
%Perets, H. B., Gualandris, A., Kupi, G., Merritt, D., Alexander, T. 2009, ApJ, 702, 884

%\bibitem[]{Hills88}
%Hills, J. G. 1988, Nat, 331, 687 

%\bibitem[]{Brown05}
%Brown, W. R.,  Geller, M. J.,  Kenyon, S. J.,  Kurtz, M. J. 2005, ApJ,  622, L33


%======

\end{thebibliography}
\end{document}